\documentstyle[epsfig]{mn}

\newif\ifAMStwofonts
\ifoldfss
  \ifCUPmtlplainloaded \else
    \NewTextAlphabet{textbfit} {cmbxti10} {}
    \NewTextAlphabet{textbfss} {cmssbx10} {}
    \NewMathAlphabet{mathbfit} {cmbxti10} {} % for math mode
    \NewMathAlphabet{mathbfss} {cmssbx10} {} %  "   "    "
  \fi
  \ifAMStwofonts
    \ifCUPmtlplainloaded \else
      \NewSymbolFont{upmath} {eurm10}
      \NewSymbolFont{AMSa} {msam10}
      \NewMathSymbol{\upi}     {0}{upmath}{19}
      \NewMathSymbol{\umu}     {0}{upmath}{16}
      \NewMathSymbol{\upartial}{0}{upmath}{40}
      \NewMathSymbol{\leqslant}{3}{AMSa}{36}
      \NewMathSymbol{\geqslant}{3}{AMSa}{3E}

       \let\le=\leqslant
       \let\ge=\geqslant
    \fi
  \fi
\fi % End of OFSS

\ifnfssone
  \newmathalphabet{\mathit}
  \addtoversion{normal}{\mathit}{cmr}{m}{it}
  \addtoversion{bold}{\mathit}{cmr}{bx}{it}
  \newmathalphabet{\mathbfit} % math mode version of \textbfit{..}
  \addtoversion{normal}{\mathbfit}{cmr}{bx}{it}
  \addtoversion{bold}{\mathbfit}{cmr}{bx}{it}
  \newmathalphabet{\mathbfss} % math mode version of \textbfss{..}
  \addtoversion{normal}{\mathbfss}{cmss}{bx}{n}
  \addtoversion{bold}{\mathbfss}{cmss}{bx}{n}
  \ifAMStwofonts
    \ifCUPmtlplainloaded \else
      \UseAMStwoboldmath
      \makeatletter
      \new@mathgroup\upmath@group
      \define@mathgroup\mv@normal\upmath@group{eur}{m}{n}
      \define@mathgroup\mv@bold\upmath@group{eur}{b}{n}
      \edef\UPM{\hexnumber\upmath@group}
      \new@mathgroup\amsa@group
      \define@mathgroup\mv@normal\amsa@group{msa}{m}{n}
      \define@mathgroup\mv@bold\amsa@group{msa}{m}{n}
      \edef\AMSa{\hexnumber\amsa@group}
      \makeatother
      \mathchardef\upi="0\UPM19
      \mathchardef\umu="0\UPM16
      \mathchardef\upartial="0\UPM40
      \mathchardef\leqslant="3\AMSa36
      \mathchardef\geqslant="3\AMSa3E

       \let\le=\leqslant
       \let\ge=\geqslant
    \fi
  \fi
\fi % End of NFSS release 1

\ifnfsstwo
  \DeclareMathAlphabet{\mathbfit}{OT1}{cmr}{bx}{it}
  \SetMathAlphabet\mathbfit{bold}{OT1}{cmr}{bx}{it}
  \DeclareMathAlphabet{\mathbfss}{OT1}{cmss}{bx}{n}
  \SetMathAlphabet\mathbfss{bold}{OT1}{cmss}{bx}{n}
  \ifAMStwofonts
    \ifCUPmtlplainloaded \else
      \DeclareSymbolFont{UPM}{U}{eur}{m}{n}
      \SetSymbolFont{UPM}{bold}{U}{eur}{b}{n}
      \DeclareSymbolFont{AMSa}{U}{msa}{m}{n}
      \DeclareMathSymbol{\upi}{0}{UPM}{"19}
      \DeclareMathSymbol{\umu}{0}{UPM}{"16}
      \DeclareMathSymbol{\upartial}{0}{UPM}{"40}
      \DeclareMathSymbol{\leqslant}{3}{AMSa}{"36}
      \DeclareMathSymbol{\geqslant}{3}{AMSa}{"3E}

       \let\le=\leqslant
       \let\ge=\geqslant
    \fi
  \fi
\fi % End of NFSS release 2

\ifCUPmtlplainloaded \else
  \ifAMStwofonts \else % If no AMS fonts
    \def\upi{\pi}
    \def\umu{\mu}
    \def\upartial{\partial}
  \fi
\fi

\title{Chemical evolution and nature of Damped Lyman-$\alpha$ systems  }
\author[F. Calura, F. Matteucci, G. Vladilo]
       {F. Calura$^{1}$\thanks{E-mail: fcalura@ts.astro.it}, 
        F. Matteucci$^{1,2}$, G. Vladilo$^{2}$ \\
        (1) Dipartimento di Astronomia-Universit\'a di Trieste, Via G. B. Tiepolo
	11, 34131 Trieste, Italy\\
	(2) INAF, Osservatorio Astronomico di Trieste, Via G. B. Tiepolo
	11, 34131 Trieste, Italy\\
	 }
	
\date{Accepted ???? .
      Received ???? ;
      in original form ????}

\pagerange{\pageref{firstpage}--\pageref{lastpage}}
\pubyear{2002}

\begin{document}

\maketitle

\label{firstpage}

\begin{abstract}We study the nature of Damped Lyman - $\alpha$ systems (DLAs)
by means of a comparison between observed abundances and models of chemical 
evolution of galaxies of different morphological type.
In particular, we compare for the first time the abundance 
ratios as functions of metallicity and redshift with dust-corrected data.
We have developed detailed models following the evolution of 
several chemical elements (H, D, He, C, N, O, Ne, Mg, Si, S, Fe, Ni and Zn)
for elliptical, spiral and irregular galaxies. Each of the models is 
calibrated to reproduce the main features of a massive elliptical, the 
Milky Way and the LMC, respectively. In addition, we run some models also
for dwarf irregular starburst galaxies. 
All the models share the same uptodate nucleosynthesis prescriptions but 
differ in their star formation histories.
The role of SNe of different type (II, Ia) is studied in each 
galaxy model together with detailed and up to date nucleosynthesis 
prescriptions.
Our main conclusions are: 1) when dust depletion is taken into account 
most of the claimed $\alpha/Fe$ overabundances disappear and DLAs show 
solar or subsolar abundance ratios.
2) The majority of DLAs can be  explained either by disks of spirals observed at 
large galactocentric distances or by irregular galaxies like the LMC
or by starburst dwarf irregulars observed at different times after the last 
burst of star formation.
3) Elliptical galaxies cannot be DLA systems since they reach a too high 
metallicity at early times and their abundance ratios show overabundances of 
$\alpha$-elements relative to Fe over a large range of [Fe/H].
4) The observed neutral gas cosmic evolution is compared with our predictions
but no firm conclusions can be drawn in the light of the available data.
\end{abstract}

\begin{keywords}
Galaxies: abundances; Galaxies: evolution; Galaxies: high redshift;
Cosmology: observations.
\end{keywords}

\section{Introduction}
The observations of QSO spectra has led to the discovery of a class of
objects, found almost along any line of sight, whose
study has induced a huge progress in the  field of galactic
evolution research. Among these objects, DLA
systems allow us to have the most direct insight into the universe as it
was soon after the growth of galactic structures.  These objects possess
a high neutral
gas content (they are characterized by the highest column density of
neutral hydrogen, $N(HI) \ge 2 \cdot 10^{20} cm^{-2}$) and 
metal abundances, which can span from $\sim 1/100$ up to $\sim 1/3$ of
the solar value.  
The
high accuracy achievable in the hydrogen column density
determinations, owing to their occupation of the damped part of the
curve of growth (Bechtold 2001), allows us to assess precise chemical
abundances for many low ionization species, such as SiII, FeII, ZnII.
Notwithstanding this, the real nature of
DLAs is far from being clear.  Initially, on the basis of kinematical
considerations,  DLAs were associated with rotating proto-disks observed
at epochs before substantial gas consumption has taken place 
(see Prochaska \&
Wolfe 1997). At the same time, the intermediate redshift imaging has
evidenced a variety of morphological types belonging to the DLA
population,  such as low surface brightness (LSB) galaxies, dwarf
galaxies and spirals (Le Brun et al. 1997).\\ In chemical evolution
models absolute abundances usually depend  on all the  model
assumptions, whereas the abundance ratios depend only  on
nucleosynthesis, stellar lifetimes and IMF. Abundance ratios can therefore
be
used as cosmic clocks if they involve two elements formed on quite
different timescales, typical examples being [$\alpha$/Fe] and [N/$\alpha$]
ratios. These ratios, when examined together with [Fe/H], or any other
metallicity tracer such as [Zn/H], allow us to clarify the
particular history of star formation involved, as shown  by
Matteucci (1991, 2001). In fact, in a regime of high star formation rate we
expect to observe overabundances of $\alpha$-elements for a large
interval  of [Fe/H], whereas the contrary is expected for a regime
of low star formation. This is due to the different roles played by  the
supernovae (SNe) of type II relative to the SNe of type Ia.  These latter
in fact, are believed to be responsible for the bulk of Fe and Fe-peak
element production and occur on  much longer timescales than SNe II,
which are responsible for the  production of the $\alpha$-elements
(i.e. O, Ne, Mg, Si, Ca).  For this reason, the analysis of the
relative abundances can enable us to have important hints on
the nature and age  of the (proto-)galaxies which give rise to DLA
systems.\\   
In the present work we compare model predictions for
chemical  abundances in galaxies of different morphological type with
DLA observed  abundances.  Therefore, our concern is to distinguish
between abundance patterns  dominated by type Ia and type II SNe,
which are usually recognizable through the lack or the presence,
respectively, of enhancement of $\alpha-$ elements relative to
iron-peak elements.  Another chemical element of interest is $^{14}$N, whose
nucleosynthetic origin is  particularly complex and uncertain. 
This element is
mainly produced by low and intermediate mass stars ($0.8 \le
M/M_{\odot} \le 8$)  with a small fraction arising from massive
stars. Nitrogen is mainly a  secondary element, in the sense that
is produced in proportion to the  carbon and oxygen originally present
in the star. This characteristic enhances the delay of production of
nitrogen owing to the delay resulting from its stellar
progenitors.  However, part of N can have a primary origin, in the
sense that it can be  produced starting from C and O manufactured by the
star ``in situ''. 
This can happen  during the Asymptotic Giant Branch (AGB) phase
when hot bottom burning  acts in  conjunction with the third dredge-up
(Renzini \& Voli, 1981).  This possibility  has been recently
confirmed also by other authors (Marigo et al.  1996;1998 and Van den
Hoeck and Groenwegen 1997).  Very recently, Maeder and Meynet (2001)
have suggested that rotation in  massive stars can be responsible for
the production of some primary N.  Thus the [N/O]
and in general the [N/$\alpha$] ratio represents another important 
diagnostic tool in chemical evolution, with N restored to the 
interstellar medium (ISM) with a  large delay relative 
to $\alpha$-elements.\\
 Previous
studies of the nature of DLA systems by means of chemical  evolution
models (Matteucci et al. 1997; Jimenez et al. 1999; Mo et al. 1998;
Prantzos \& Boissier 2000) have  suggested that  some of these objects
can be disks of spirals in formation, whereas others can be low surface
brightness objects and even starbursting dwarfs similar to the local
very metal poor star-bursting galaxy IZw18.
However, the abundance patterns observable in DLA systems can be interpreted 
by means of chemical evolution models only if they are driven
by pure nucleosynthesis. 
The presence of dust in DLAs (which is indicated by several clues such 
as the reddening of QSOs in presence of DLAs along the line of sight, see Fall
and Pei 1995) can represent a serious complication in pursuing 
this task, since its effect is to  
deplete some chemical elements (e. g. Fe, Si) more than others (O, Zn) 
thus altering the observed abundances. 
For this reason, in this paper we have applied dust-corrections to 
a set of measured data according to the dust model 
discussed in Vladilo (2002a,b), 
and we have re-interpreted the observations, reaching 
conclusions very different from the ones drawn without considering dust
contamination.

The paper is organized as follows: in section 2 we will describe 
the current observational and theoretical status 
of the study of DLA systems; in section 3 we will
present a description of the adopted chemical evolution models;
in section 4 we will discuss our results and draw the
conclusions in section 5.
Throughout the paper we express chemical abundances normalized 
to the solar abundance
values measured by Anders and Grevesse (1989)
($[X/H]=log(X/H)-log(X/H)_{\odot}$).

\section[]{Observational and theoretical background}
\subsection {DLA observations}
The strong absorption lines originating in DLA systems can usually be 
detected with moderate resolution, and the damping profile of the
lines permits confortable and accurate measurements of the neutral gas 
column density, which is crucial in chemical abundance studies
(Bechtold 2001).
For this reason DLA systems are particularly suitable to chemical 
evolution studies of the high-redshift universe, where they 
can be found in large number and allow us to observe galaxies as they 
were in the very early phases of their history.
So far, chemical abundances have been measured in a large ($\approx 180$, 
Molaro 2001) sample of known DLAs (see e. g. 
Pettini et al. 1994, Pettini et al. 1995, Pettini et al. 1997, Lu et al. 
1996, Lu et al. 1998, Prochaska \& Wolfe 1999, Centuri\'on et al. 2000, 
Prochaska \& Wolfe 2002 
and references therein). The major concerns in abundance
determinations are represented by
dust-depletion and ionization correction effects.
Usually ionization correction effects are neglected because
the abundances are derived from dominant ionization states of 
the elements of
interest.
In fact, detailed photoionization computations indicate that ionization 
corrections are usually negligible for the most common elements investigated
so far (Howk et al. 1999, Izotov et al. 2001, Vladilo et al. 2001).
In the present work particular attention is paid to dust-depletions 
effects, in particular to the variation of dust-corrections depending
on changes in the chemical composition of the DLA systems.

\subsubsection {Dust depletions}
There are several indications about the presence of dust within DLAs, 
such as reddening of background quasars in presence of DLA absorbers
(Fall \& Pei 1995), or gas-phase
abundances of trace elements such as Cr, Zn (Pettini et al. 1994).  
Observations provide hints for selective depletions acting in dense clouds
within our Galaxy (Jenkins 1987).
The dust content 
in the highest-redshift systems seems to be reduced relative to the
interstellar clouds in our Galaxy 
(Pettini et al. 1997), possibly owing 
to the efficient grain-disruption in interstellar 
shocks following bursts of star formation (Pettini \& Bowen 1997).
Several authors have 
developed a formalism aimed at quantifying the effect of dust depletion 
on DLA abundances by assuming that the dust in these objects is similar to the
Galactic interstellar dust and scaling 
the amount of depletion by including variations of the
dust-to-metal ratios (Kulkarni et al. 1997, Vladilo 1998).
These methods did not take into account variations of the chemical 
composition of dust among different systems.
Recently, Vladilo (2002a,b) derived an analytical expression 
which takes into account the interstellar 
depletions according to the variations of the physical conditions and chemical 
abundances of the medium. 
From this scaling law, a new method for estimating dust depletion 
corrections in DLAs as a function of a set interstellar parameters has been 
presented, together with a preliminary sample of dust corrected abundances.
In the present work we compare, for the first time, dust corrected abundances
obtained using this new method with detailed predictions of chemical evolution
models.
%The stability of the corrected abundances against variations of the input 
%parameters of the dust correction procedure is also considered (DOVE).

\subsubsection {Are DLAs really $\alpha-$ enhanced?}
%The [$\alpha$/Fe] ratio is of particular importance  in DLA studies
%because of the different nucleosynthetic origin  of the
%$\alpha$-elements and $Fe$-peak elements, being the former produced
%mainly by massive stars and restored into the ISM through type II SN
%explosions whereas the latter being formed during type Ia SNe explosions,
%therefore contributing to the ISM enrichment on longer timescales.
The [Si/Fe] ratio can be easily measured in DLA systems.
Its interpretation
is however difficult owing to the differential effect that dust has on
the abundances of Si and Fe.
Several authors have pointed out that the
[$\alpha$/Fe] ratios in DLAs are consistent with the abundances measured
in metal-poor stars of the Milky Way (Lu
et al. 1996, Prochaska et al.  2001a; Pettini et al. 2000),
i. e. with a chemical evolution pattern dominated by type II SN.  The
mean observed value gives in fact $[Si/Fe]=+0.39 \pm 0.17$ (Molaro
2001), which is similar to the value observed in Galactic halo stars
with metallicities comparable to the values measured in DLAs.
The observed enhancement, if real and not affected by dust depletion, 
could be associated with the halo phase of 
a forming disk or
to the formation of a spheroid.
However, in other cases the [Si/Fe] measurements  show 
little sign of $\alpha$- enhancement  (Ellison et al. 2001, Pettini et
al. 1999).
Pettini et al. (1999) consider a set of absorbers at low (z $<1.5$) 
redshift and find no evidence
for an enhancement of $\alpha$-elements over iron 
and
argue that in LSB galaxies or in 
the outskirts of a disk, 
where star formation proceeds more
slowly, we could observe nearly solar [$\alpha$/Fe] values also 
at low metallicities.
So, in this case, the lack of a clear $\alpha$ enhancement would 
not necessarily exclude a
connection between DLA sites and galactic disks, 
provided that we are
observing the external regions.
As is suggested in Vladilo (1999), the fact that the differential 
cross-section for DLA absorption is
$dA\simeq 2\pi r dr$ (for a disk seen face-on) implies that galactic 
regions with larger $r$
have
higher probabilities of detection.\\ 
The [S/Zn] ratio represents a valid 
alternative and nevertheless a more reliable diagnostic, since both
elements are unaffected by dust depletion. 
However, the major concern in estimating S abundances is represented by
the Ly$\alpha$ forest,
whose contamination can have serious effects on the measurements.
The available [S/Zn] data do not show the typical enhancement 
observed in Galactic metal-poor
stars at metallicity levels comparable to the ones of DLAs 
(Centuri\'on et al. 2000).
The emerging picture seems to point
towards different chemical evolution patterns,  characterized either by
halo-like or solar abundance ratios. This could be consistent either
with a heterogeneous population of  progenitors or with a single type
of objects observed at various phases  of their evolution.

\subsubsection {Metallicity evolution of DLAs}
The metallicity in DLA systems is usually traced by Zn or Fe, whose
abundances are easily  measurable in DLAs.  Metallicities in solar
units span the interval $1/100$-$1/3$ solar,
with only a modest increase of the metallicity with decreasing
redshift (Pettini et al. 1997, Vladilo et al. 2000, Prochaska \& Wolfe 2001).
Once the column density weighted metallicity is
considered and the diagram is interpreted as an evolutionary one, then 
there is no statistically  significant sign of evolution of the [Zn/H] 
abundance with redshift in the sense that the systems at lower redshift 
are the most evolved ones.
(Pettini et al. 1999, Vladilo et al. 2000, Prochaska
et al. 2001).  
However, it is dangerous to interpret the [Zn/H] vs z diagram in terms 
of an age-metallicity relation, since it could rather reflect
the specific evolutionary state of different objects formed at different
cosmic times, and not necessarily different timesteps in the evolution 
of objects formed at the same time.  
Moreover,
in interpreting the behaviour of the DLA metallicity, 
one should keep in mind that 
the sample of DLAs may be biased against detection of systems of high metal 
column density. A hint in this direction comes from the results by 
Boiss\'e et al. (1998),
who traced a [Zn/H] vs log N(HI) diagram and noticed an apparent
anti-correlation between the metallicity and the column density which
is  physically difficult to interpret.  This could be the result of an
observational bias against high metallicity, high density  systems
which would obscure the QSOs behind them. This fact would imply that 
these objects
can be underepresented in magnitude-limited  samples
(Pettini et al.  1999, Molaro 2001), and could even cause a high
incidence of non-spiral morphologies in the samples (Boiss\'e et
al. 1998).  The absence of high metallicity, high N(HI) systems could
alternatively be explained by higher  metallicity regions having
significantly consumed their gas reservoirs and therefore having lower HI
column densities, as suggested by  Prochaska \& Wolfe (2001).

\subsection {DLA models}
Several models of DLAs have been developed so far with various techniques 
in order 
to reproduce their observed properties, 
usually encountering most difficulties in explaining the mild evolution 
in the metallicity (Prochaska et al. 2001a).

Pei and Fall (1995) treated chemical evolution with a new approach: 
they considered 
large comoving volumes, representative 
of the universe as a whole and containing many DLA systems, and studied 
the 
metallicity evolution inside these spaces assuming
instantaneous recycling approximation
in the cases of closed-box, inflow and outflow models.
They incorporated in their models a self-consistent correction 
for the absorbers 
that are missing from the existing samples, which drop out
as a result of dust-obscuration of background quasars, assuming a 
dust-to-gas ratio 
proportional to the mean metallicity.
Their result reproduced the average properties of the universe 
in the past and at the present epoch, but cannot make any prediction on which
morphological types sould be included or excluded from the DLA current 
samples.\\
Matteucci et al. (1997) studied the relative abundances observed in 
several DLA systems; 
in particular they studied the $\alpha$/Fe
and N/O ratios by means of chemical evolution models for starburst galaxies 
and the solar neighbourhood.
They suggested that the high (N/O) ratios observed in a couple of DLAs 
(the ones towards QSO0000-26 and QSO1331+1700)
indicate a different chemical evolution than in the Milky Way and spirals in
general, and demonstrated that the most promising models to explain 
the observed abundances are those applied to dwarf
irregular galaxies. However, they  concluded that the differences in the 
abundances observed among different DLAs could be due to
a morphologically heterogeneous population. 

Cen and Ostriker (1999) used numerical simulations to study the 
metallicity evolution of regions of the universe characterized by
different overdensities $\delta_{\rho}$, identifying DLAs with zones 
with $\delta_{\rho}\sim 10^{2}$.
They found that the median metallicity in these portions of the universe 
increases only slightly from $z\sim3$ to $z\sim0.5$, with a large
metallicity spread expected at any redshift, in agreement with observations. 
They  concluded that DLAs do not represent the highest
metallicity sites at any time.\\ 
An attempt to identify the morphology of DLAs and their 
compatibility with galactic 
disks was made by Mo, Mao and White (1998), who studied
the population of galactic disks in hierarchical clustering models 
for galaxy formation. 
They concluded that galaxies selected as DLAs should be biased
towards low surface densities, pointing to LSB galaxies as 
the main candidates 
for the high-redshift DLA population.\\
Jimenez et al. (1999) coupled a disk formation model to a chemical evolution 
model to explore the metallicities, dust content 
and gas mass density of LSB and high surface brightness (HSB) galaxies; 
both high-z 
LSB and low-z HSB can reproduce the 
observed metallicities, but HSB fail in reproducing the gas density 
evolution, 
so they suggested that LSB disks could dominate 
the DLA high-z population.\\
Prantzos \& Boissier (2000) used chemical evolution models for disk 
evolution and found that the mild evolution in the column-density
weighted metallicity is consistent with galactic disks and showed how 
observational biases (such as reddening of the background QSOs by the
most dense and metal rich systems) can be consistent with the no-evolution 
picture emerging from the current observations.

The global picture emerging from these studies converges 
to the result that in general a 
heterogeneous DLA population can 
explain the observations better than a homogeneous one.

\section{Model description}
In the present work the chemical properties of DLA systems are investigated
by means of detailed chemical evolution models of galaxies of
three different morphological types: ellipticals, spirals and irregulars.
These models allow one to follow in detail the evolution of the abundances 
of several chemical
species, starting from the matter reprocessed 
by the stars and restored into the ISM through stellar winds and supernova
explosions. 
Detailed descriptions of the chemical evolution models can be found in 
Matteucci
and Tornamb\'{e} (1987) and Matteucci (1994) for elliptical galaxies, 
Chiappini et al. (1997, 2001) for the spirals,   
Bradamante et al. (1998) and Recchi (2002) for irregular galaxies.
All of these models have been updated to include the same nucleosynthesis 
prescriptions (see later).
According to our scheme, elliptical galaxies form as the result of the rapid 
collapse of a homogeneous sphere of
primordial gas (Matteucci 1994) where star formation is taking place at 
the same time 
as the collapse proceeds, and evolve as
"closed-boxes", i.e. without any interaction with the external environment. 
Star formation is assumed to stop when  the energy of the ISM, 
heated by stellar 
winds and SNe explosions,
is equal to the binding energy of the gas. 
At this time a galactic wind occurs, 
sweeping away all 
the residual gas in the galaxy. 

Spiral galaxies are assumed to form as a result of two main infall episodes
(Chiappini et al. 1997). During the first episode the halo forms and 
the gas shed by the 
halo rapidly
gathers in the center leading to the formation of the bulge. 
During the second episode, 
a slower infall
of external gas gives rise to the disk with the gas 
accumulating faster in the inner than 
in the outer
region ("inside-out" scenario, Matteucci \& Fran\c cois, 1989). The process of disk formation is 
much longer than the halo
and bulge formation, with time scales varying from $\sim2Gyr$ in the 
inner disk, 
$\sim8 Gyr$ in the solar region, and  up to $15-20
Gyr$ in the outer disk.\\    
Irregular galaxies are assumed to form owing to a continuous infall 
of pristine gas, until a mass of
$\sim 10^{9}M_{\odot}$ is accumulated.
The star formation rate proceeds either in bursts 
separated by
long quiescent periods or at a low regime but continuously.
We have computed also a new model to describe the 
Magellanic irregulars; this 
model is tuned to reproduce the Large Magellanic Cloud (LMC), and assumes 
two main bursts of star formation with low but continuous star formation 
in between.

The main features of the models used here are the following:\\
1) one-zone for irregulars and ellipticals, multi-zone for spirals 
with instantaneous 
and complete mixing of gas in each zone;\\
2) no instantaneous recycling approximation (the stellar lifetimes are properly taken 
into account);\\
3) the evolution of several chemical elements  (H, D, He, C, N, O, Ne, Mg, Si, S, Fe, Ni and Zn)
is followed in detail.\\
4) The nucleosynthesis prescriptions are common to all the models and 
include: the yields 
of Thielemann, Nomoto \& Hashimoto (1996) for massive stars
(M $> 10 M_{\odot}$), 
the yields of van den Hoeck \& Groenewegen (1997) for
low and intermediate mass stars ($0.8 \le M/M_{\odot} \le 8$) and the yields of 
Nomoto et al. (1997) for type I a SNe. For the evolution of Zn and Ni we adoped
the nucleosynthesis prescriptions of Matteucci et al. (1993).

\subsection{The theory}

Let $G_{i}$ be the fractional mass of the element $i$ in the gas
within a galaxy, its time-evolution is described by the basic equation:
 
\begin{equation}
\dot{G_{i}}=-\psi(t)X_{i}(t) + R_{i}(t) + (\dot{G_{i}})_{inf} - 
(\dot{G_{i}})_{out}
\end{equation}   

where $G_{i}(t)=M_{g}(t)X_{i}(t)/M_{tot}$ is the gas mass in the form of an
element $i$ normalized to a total initial mass $M_{tot}$. 
$M_{gas}$ and $M_{tot}$ are substituted by $\sigma_{gas}$ and $\sigma_{tot}$ 
in the case of galaxy disks.
The quantity 
$X_{i}(t)=
G_{i}(t)/G(t)$ represents the fractional abundance by  mass of an 
element $i$, with
the summation over all elements in the gas mixture being equal to unity.
$\psi(t)$ is the star formation rate (SFR), the fractional 
amount
of gas turning into stars per unit time. $R_{i}(t)$ represents the returned
fraction of matter in the form of an element $i$ that the stars eject into 
the 
ISM through stellar winds and 
supernova explosions; this term contains all the prescriptions regarding the stellar yields and
the SN progenitor models.
The two terms 
$(\dot{G_{i}})_{inf}$ and  $(\dot{G_{i}})_{out}$ account for the infall
of external gas and for galactic winds,
respectively.
The quantity $G(t)= M_{g}(t)/M_{tot}$ is the total fractional mass of gas
present in the galaxy at the time t.
The main feature characterizing a particular morphological galactic type is
represented by the prescription adopted for the star formation history.

In the cases of ellipticals and irregular
galaxies the SFR $\psi(t)$ (in $Gyr^{-1}$) has a simple form and is given by:

\begin{equation}
\psi(t) = \nu G(t) 
\end{equation}
The quantity $\nu$ is the efficiency of star formation, namely the inverse of
the typical time scale for star formation. 

In the case of ellipticals, 
$\nu \sim 11 Gyr^{-1}$ (for a baryonic mass of $10^{11} M_{\odot}$) is
assumed, and it drops to zero at the onset of a galactic wind, 
which develops as the
thermal energy of the gas heated by supernova explosions exceeds the binding
energy of the gas (Matteucci and Tornamb\'{e} 1987),
and rids the galaxy of all the residual gas. 
This quantity is strongly
influenced by assumptions concerning the presence and distribution of dark
matter (Matteucci 1992); for the model adopted here a diffuse 
($R_e/R_d$=0.1, where
$R_e$ is the effective radius of the galaxy and $R_d$ is the radius 
of the dark matter core) but 
massive ($M_{dark}/M_{Lum}=10$) dark halo has 
been assumed (see Bertin et al. 1992).

In the case of Magellanic irregular galaxies, we have considered a 
burst-like star formation history (see Bradamante et al. 1998)
which is successful in reproducing the [O/Fe] vs [Fe/H] data observed 
in the Large Magellanic Cloud (LMC) (Hill et al. 2000)
and the final LMC gas fraction.
Figure 1 shows the evolution in time of the adopted star formation 
rate (upper panel) and the comparison 
between the predictions
and the observations for the [O/Fe] vs [Fe/H] (lower panel).
The star formation is characterized by two bursts, 
the first occurring at $\sim 2 Gyrs$ and the second at $\sim 13 Gyrs$, 
both of them with a duration of $2 Gyr$. During the period in-between 
the two bursts, a low and constant star formation
is taking place. The efficiency of star formation in this case is 
$\nu =0.1 Gyr^{-1}$. As can be seen from the figure, 
such a star formation history reproduces satisfactorily 
the observed [O/Fe] vs [Fe/H] pattern. 
The final gas fraction is, according to our model, $f_{gas}=0.17$ 
($f_{gas}=M_{gas}/M_{tot}$), in agreement
with the value measured by Lequeux et al (1979, $f_{gas}= 0.15$).
The possibility of a galactic wind is taken into account also for 
these systems always under the same conditions as for ellipticals. 
In this case, the wind normally develops during the starburst and 
ceases immediately afterwards. The wind rate is assumed to be proportional 
to the star formation rate. 
The prescriptions for the dark matter halo are similar to those
adopted for the ellipticals,
namely a massive but diffuse dark halo 
(see Bradamante et al. 1998 for details).
%%%%%%%%%%%%%%%%%%%%%%%%%%%%%%%%%%%%%%%%%%%%%%%%%%%%%%%%%%%%%%%%%%%%%%%%%%%%%%%%%%%%%%%%%%%%%%%%%%%%%%%%%%%%%%%%%%
\begin{figure*}
\centering
\vspace{0.001cm}
\epsfig{file=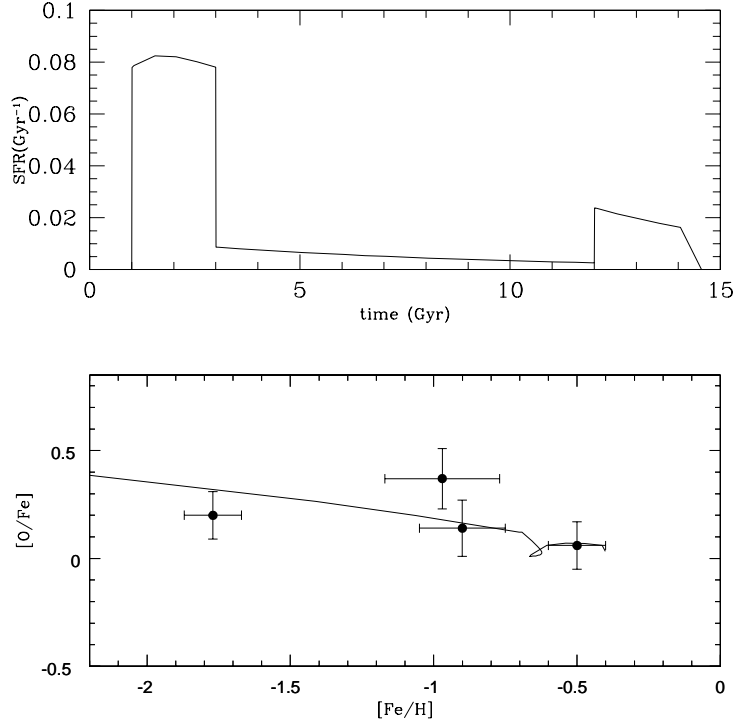,height=10cm,width=10cm}
\caption[]{Adopted star formation rate (above) for the irregular model
and comparison between the predicted [O/Fe] vs [Fe/H] evolution
and the values observed in LMC stars (below). Data taken from Hill et al. (2000).
}	
\end{figure*}
%%%%%%%%%%%%%%%%%%%%%%%%%%%%%%%%%%%%%%%%%%%%%%%%%%%%%%%%%%%%%%%%%%%%%%%%%%%%%%%%%%%%%%%%%%%%%%%%%%%%%%%%%%%%

In the case of spiral galaxies, the SFR expression (Chiappini et al. 1997) is:

\begin{equation}
\psi(r,t) = \nu [\frac{\sigma(r,t)}{\sigma(r_{\odot},t)}]^{2(k-1)} [\frac{\sigma(r,t_{Gal})}{\sigma(r,t)}]^{k-1}
\sigma^{k}_{gas}(r,t)
\end{equation}
where $\nu$ is the star formation efficiency, $\sigma(r,t)$ is the total mass 
(gas + stars) surface density at a radius r and time t,
$\sigma(r_{\odot},t)$ is the total mass surface density in the solar 
region and $\sigma_{gas}(r,t)$ is the gas surface mass
density. For the gas density exponent $k$ a value of 1.5 has been assumed 
by Chiappini et al. (1997) in order to ensure a good
fit to the observational constraints in the solar vicinity.
The efficiency of star formation is set to $\nu=1 Gyr^{-1}$, and
becomes zero when the gas surface density drops below a certain 
critical threshold; we adopt a threshold
density $\sigma_{th}\sim 7 M_{\odot} pc^{-2}$ in the disk as suggested by 
Chiappini et al. (1997).
Such value corresponds to a column density of $N_{HI} \sim 8.35 \cdot 10^{20} cm^{-2}$,
i. e. well above the DLA detection limit of $2 \cdot 10^{20} cm^{-2}$. 

The spiral disk is approximated by several independent rings, 2 kpc wide, 
without exchange of matter between them. 

The rate of accretion of matter, which is assumed to have 
primordial composition, 
in each shell is:

\begin{equation}
\frac{d\sigma_{I}(r,t)}{dt}=A(r) e^{-t/\tau_{H}} + B(r)e^{-(t-t_{max})/\tau_{D}(r)}
\end{equation}   
where $\sigma_{I}(r,t)$ is the surface mass density of the infalling 
pristine gas; 
$t_{max}$ is the time of maximum gas accretion onto
the disk, coincident with the halo/thick-disk phase and set equal to 1 Gyr; 
$\tau_{H}$ and $\tau_{D}(r)$ are the timescales for
mass accretion onto the halo/thick-disk and thin-disk components respectively.
In particular, the timescale for the formation of the inner halo,
$\tau_{H}=0.8 Gyr$,  is assumed to be constant whereas the timescale for 
the formation of the disk is a function of the galactocentric distance, 
$\tau_{D}(r) = 1.033 \times r -1.267 Gyr$, 
according to the "inside-out" scenario (Chiappini et al. 1997).
The quantities $A(r)$ and $B(r)$ are derived from the condition 
for reproducing the 
current total mass surface density distribution
in the halo and along the disk, respectively, at the present time.

The initial mass function (IMF) is usually assumed to be constant in space and
time in all the models and is expressed by the formula:
 
\begin{equation}
\phi(m) = \phi_{0} m^{-(1+x)} 
\end{equation}

where $\phi_{0}$ is a normalization constant.
In the case of irregulars and ellipticals a Salpeter-like IMF (1955) 
($x=1.35$) 
has been assumed, 
whereas for spirals the Scalo (1986) prescription has been adopted
($x=1.35$ for $0.1 \le m/m_{\odot} \le 2$, $x=1.7$ for $2 < m/m_{\odot} 
\le 100$). 
These choices of the IMF have been demonstrated to be the best in reproducing 
the observational constraints
in the solar neighbourood 
and in the nearby galaxies.   
In all cases the assumed mass range is
$0.1-100 M_{\odot}$.  

Tables 1, 2, 3 summarize the adopted model parameters for ellipticals, 
spirals and irregulars, respectively.
%%%%%%%%%%%%%%%%%%%%%%%%%%%%%%%%%%%%%%%%%%%%%%%%%%%%%%%%%%%%%%%%%%%%%
\begin{table*}
\begin{flushleft}
\caption[]{Model parameters for a typical elliptical galaxy. $M_{tot}$ is the 
baryonic mass of the galaxy, $R_{e}$ is the effective radius, 
$\nu$ is the star-formation efficiency.}
\begin{tabular}{l|llll}
\noalign{\smallskip}
\hline
\hline
\noalign{\smallskip}
Elliptical & $M_{tot} (M_{\odot})$& $R_{e}(kpc)$ & $\nu(Gyr^{-1})$ & $IMF$ \\
\noalign{\smallskip}
\hline
\noalign{\smallskip}
 &$10^{11}$ &3&11.2&Salpeter\\
%\noalign{\smallskip}
\hline
%\noalign{\smallskip}
\hline
\end{tabular}
\end{flushleft}
\end{table*}
%%%%%%%%%%%%%%%%%%%%%%%%%%%%%%%%%%%%%%%%%%%%%%%%%%%%%%%%%%%%%%%%%%%%%%%%%%%%%%%%
%%%%%%%%%%%%%%%%%%%%%%%%%%%%%%%%%%%%%%%%%%%%%%%%%%%%%%%%%%%%%%%%%%%%%
\begin{table*}
\vspace{0cm}
\begin{flushleft}
\caption[]{Model parameters for a typical spiral galaxy. $M_{tot}$ is the baryonic mass of the galaxy, 
$\tau_{H}$ is the infall timescale for the
halo, 
$\nu$ is the star-formation efficiency.}
\begin{tabular}{l|llll}
\noalign{\smallskip}
\hline
\hline
\noalign{\smallskip}
Spiral & $M_{tot} (M_{\odot})$& $\tau_{H}(Gyr)$ & $\nu(Gyr^{-1})$ & $IMF$ \\
\noalign{\smallskip}
\hline
\noalign{\smallskip}
 &$10^{11}$ &0.8&1.0&Scalo\\
%\noalign{\smallskip}
\hline
%\noalign{\smallskip}
\hline
\end{tabular}
\end{flushleft}
\end{table*}
%%%%%%%%%%%%%%%%%%%%%%%%%%%%%%%%%%%%%%%%%%%%%%%%%%%%%%%%%%%%%%%%%%%%%%%%%%%%%%%%
%%%%%%%%%%%%%%%%%%%%%%%%%%%%%%%%%%%%%%%%%%%%%%%%%%%%%%%%%%%%%%%%%%%%%
\begin{table*}
\vspace{0 cm}
\begin{flushleft}
\caption[]{Model parameters for a Magellanic Irregular galaxy. $M_{tot}$ is the mass of the galaxy, $Wind$ is the wind parameter, 
$\nu$ is the star-formation efficiency.}
\begin{tabular}{l|llll}
\noalign{\smallskip}
\hline
\hline
\noalign{\smallskip}
Irregular & $M_{tot} (M_{\odot})$& $Wind$ & $\nu(Gyr^{-1})$ & $IMF$ \\
\noalign{\smallskip}
\hline
\noalign{\smallskip}
 &$6 \cdot 10^{9}$ &0.25&1&Salpeter\\
%\noalign{\smallskip}
\hline
%\noalign{\smallskip}
\hline
\end{tabular}
\end{flushleft}
\end{table*}
%%%%%%%%%%%%%%%%%%%%%%%%%%%%%%%%%%%%%%%%%%%%%%%%%%%%%%%%%%%%%%%%%%%%%%%%%%%%%%%%

\section{Results}
\subsection{The metallicity evolution}
Some elements, such as Zn, appear essentially
undepleted in the Milky Way ISM, 
therefore they can be 
considered as reliable metallicity tracers.
The nucleosynthesis of Zn is a debated issue, since according to 
theoretical models
its production can ensue via s-processes in low/high mass stars during 
He-core burning as well as during explosive nucleosynthesis events
occurring in type Ia and II supernovae (see Matteucci et al. 1993 and 
references therein).
In particular, in our models we assume a Fe-like scaling Zn production 
on the basis 
of the results of Matteucci et al. (1993), in which a good fit to the
solar abundance of Zn is found if the bulk of its production is 
ascribed to type Ia supernovae.
On the other hand, since a non negligible fraction of iron 
comes from type II SNe (Thielemann et al. 1996) and the abundance of 
Fe seems to vary in lockstep with that of Zn in the solar vicinity, the 
SNe II should
also produce some Zn. 
In our models the amount of Zn produced through explosive nucleosynthesis 
in massive stars
scales with the Fe yield according to:\\
\begin{equation}
Y_{Zn}=\beta  \times Y_{Fe} \\
\end{equation}
where $\beta$ represents a multiplicative factor. We have run a chemical 
evolution model for the solar vicinity 
varying $\beta$ in order to reproduce the 
[Zn/Fe] vs [Fe/H] observed in Galactic stars of different metallicities 
by various authors
(see Umeda \& Nomoto 2002 and references therein).
Figure 2 shows the observational [Zn/Fe] vs [Fe/H] distribution compared 
with the predictions
for the solar neighbourhood model when two values of $\beta$ are adopted
(dashed line: $\beta=0.002$; solid line: $\beta=0.001$).
As can be seen from the figure, the role of type II SNe is important in 
determining the shape of the theoretical curve.
Although the data show a remarkable dispersion,
we have assumed the value $\beta=0.002$ since with such a prescription the 
observed pattern seems best reproduced.
For Zn produced in type Ia SN, we assume a constant value
$M_{Zn} \sim 3.2 \cdot 10^{-4} M_{\odot}$. \\
%%%%%%%%%%%%%%%%%%%%%%%%%%%%%%%%%%%%%%%%%%%%%%%%%%%%%%%%%%%%%%%%%%%%%%%%%%%%%%%%%%%%%%%%%%%%%%%%%%%%%%%%%%%%%%%%%%
\begin{figure*}
\centering
\vspace{0.001cm}
\epsfig{file=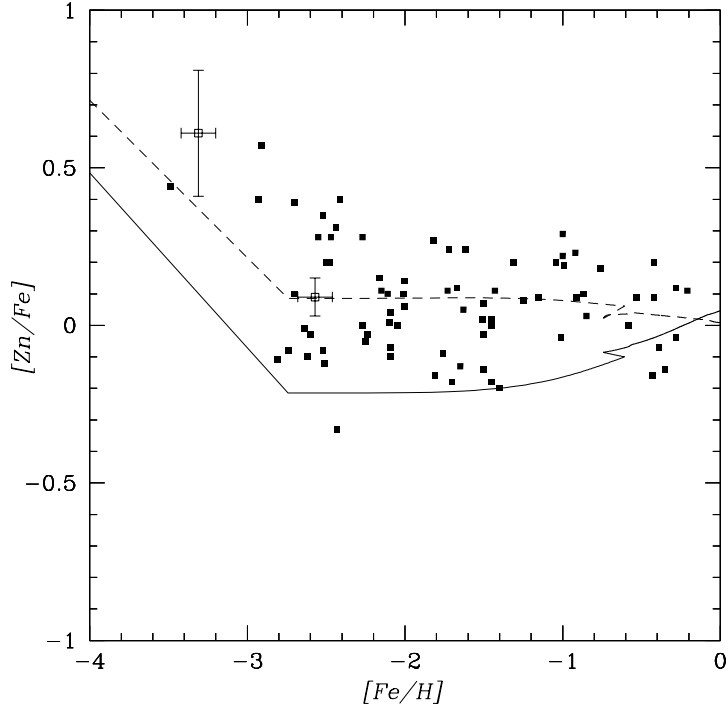,height=10cm,width=10cm}
\caption[]{Observed and predicted [Zn/Fe] vs [Fe/H] distribution in the solar neighbourhood for two different
nucleosynthetic prescriptions for the production of Zn (see text for details). \emph{Dashed line}: model with 
$\beta=0.002$; \emph{solid line}: model $\beta=0.001$.
Compilation of data taken from Umeda \& Nomoto (2002).
}	
\end{figure*}
%%%%%%%%%%%%%%%%%%%%%%%%%%%%%%%%%%%%%%%%%%%%%%%%%%%%%%%%%%%%%%%%%%%%%%%%%%%%%%%%%%%%%%%%%%%%%%%%%%%%%%%%%%%%
Figure 3 shows the evolution of the [Zn/H] ratio versus redshift for galaxies 
of different morphological types, compared to the
values observed in DLAs by various authors.
The scatter in the observed data is large and in principle 
can be due to several effects, such as different object ages,
different lines of sight as well as different galactic 
morphologies (Vladilo et al. 2000).
We consider the chemical evolution models for galaxies of 
different morphological types, i.e. spirals, ellipticals
and irregulars, as described before, and study the [Zn/H] ratio as a 
function of redshift 
assuming a given cosmology 
(the standard $\Lambda$CDM cosmology characterized by 
$\Omega_{m}=0.3, \Omega_{\Lambda}=0.7$ and $h=0.65$)
and for the sake of simplicity the same formation
redshift for all galaxies ($z_f=3$ and $5$).
In reality, different galaxies are likely to have started to form stars at different 
cosmic times. In this case, our model predictions would not change but just shift along the redshift axis, according to the adopted z-t relation.
In particular, for spiral disks we study the chemical evolution at several
galactic radii (4kpc, 8kpc, 14kpc, 18kpc). The chemical enrichment at 
different radii is different owing to the variation of the SFR 
and infall rate 
with galactocentric distance. This is a consequence of the assumed 
``inside-out'' 
picture for the formation of a galactic disk.
In Figure 3 we show the metallicity 
evolution for all the 
studied galaxies for galaxy formation redshifts $z_{f}=3$ (upper panel)
and $z_{f}=5$ (lower panel).
In the first case we have difficulties in reproducing the 
highest-redshift data because of a too late galaxy formation,
which leads to very low metallicity systems at high z (i. e. $z>2.5$),
especially
in the case of spirals at various radii and irregulars.
On the other hand, elliptical galaxies produce too high metallicities. 
%%%%%%%%%%%%%%%%%%%%%%%%%%%%%%%%%%%%%%%%%%%%%%%%%%%%%%%%%%%%
%Another important issue is the fact that at
%$z\sim3$ we already observe several systems which have 
%undergone a relevant metal enrichment.
%Furthermore, as the data show, there seems 
%to be a physical lower limit to the metallicity 
%of the DLA systems observed at 
%all epochs (see also Prochaska et al. 2001a). In fact, if lower 
%metallicities were present, there should be no 
%physical reason why they could not be detected. 
%Other data obtained at very high redshifts ($z>4$, see Molaro 2001) 
%show rather high metallicities
%(for instance   
%DLAs observed at $z_{abs}=4.446$ towards APM BR J0307-4945, with 
%$[Fe/H]=-1.97\pm 0.19$, Dessauges-Zavadsky et al 2000).
%This can be interpreted as evidence of a prompt enrichment 
%in metallicity which occurred in a very short time,
%followed by a long period characterized by a lack of noticeable 
%evolution (Molaro 2001).
%%%%%%%%%%%%%%%%%%%%%%%%%%%%%%%%%%%%%%%%%%%%%%%%%%%%%%%%%%%%%%%%%%%
In spite of the spread in the data and the fact that some 
high-redshift data are lower limits, all these issues could set
a solid constraint on the DLA formation epoch if we assume 
that the most metal poor systems represent the most
unevolved ones. 
Assuming a global event of galaxy formation 
at $z=3$, all the $z>3$ data are out of the
range spanned by the spirals at all radii. 
On the other hand, ellipticals are unlikely to be the dominating 
galaxies in the DLA population 
because of the high metal content and high [$\alpha$/Fe] ratios
predicted for these objects 
at high redshift.
If we assume instead the formation epoch at $z_f=5$, practically all data fall 
in the range of existence of the spirals at
various radii (only the 3 highest metallicity systems are excluded) 
and of the irregular galaxies. 
This leads us to suggest that the formation
epoch for the bulk of the DLA population could have occurred at 
redshifts between $z_{f}=5$ and $z_{f}=3$.
%Furthermore, 
%the comparison between data and models suggests that many DLA systems could be galactic disks observed at different galactocentric distances,
%since the large observed spread appears fairly accounted by considering a multizone model for spiral galaxies. 
\\

%%%%%%%%%%%%%%%%%%%%%%%%%%%%%%%%%%%%%%%%%%%%%%%%%%%%%%%%%%%%%%%%%%%%%%%%%%%%%%%%%%%%%%%%%%%%%%%%%%%%%%%%%%%%%%%%%%
\begin{figure*}
\centering
\vspace{0.001cm}
\epsfig{file=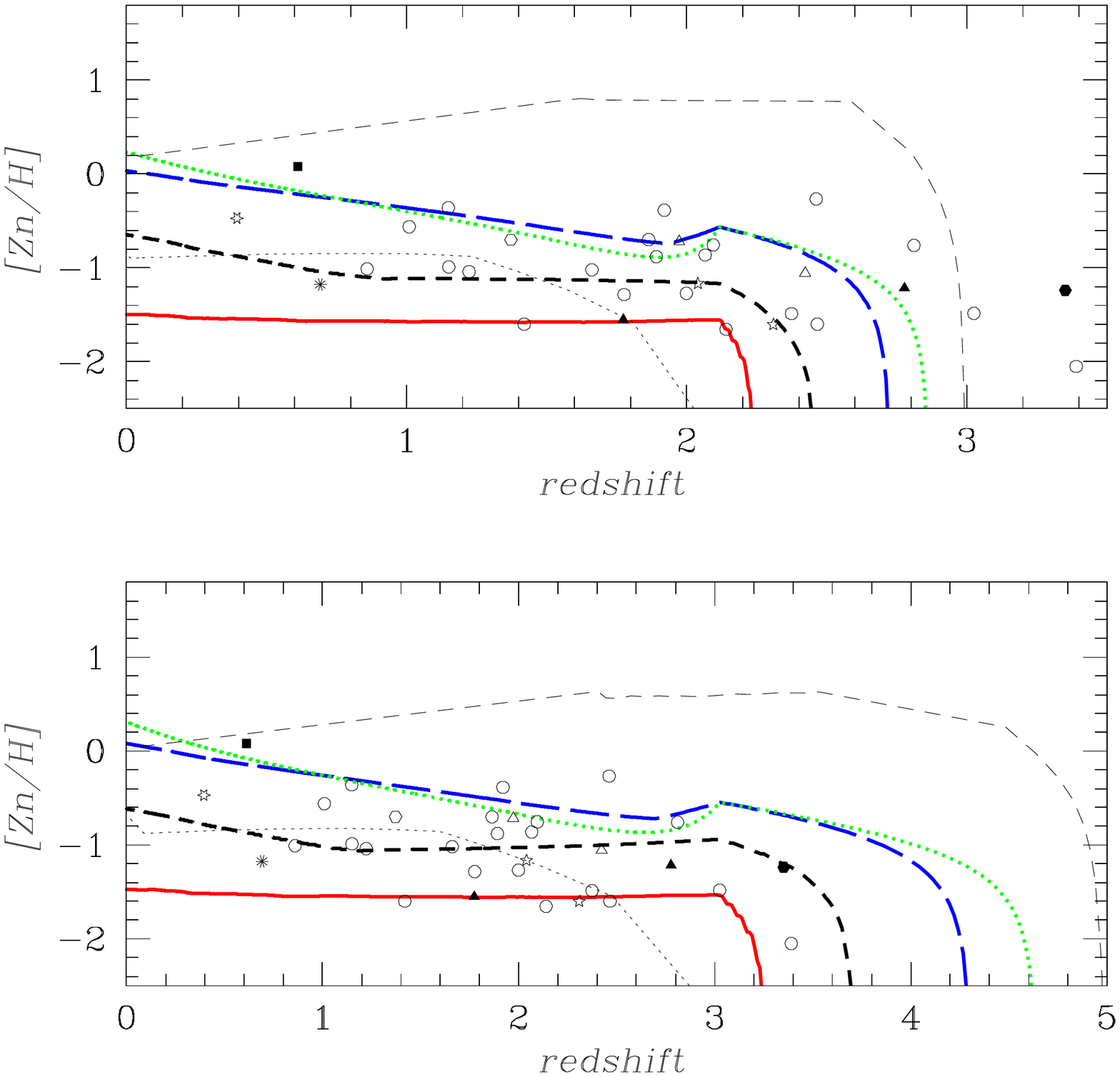,height=10cm,width=10cm}
\caption[]{Redshift evolution of [Zn/H] for galaxies of different morphological types in the case of galaxy
formation occurred at $z_{f}=3$ (upper panel) and $z_{f}=5$ (bottom panel).
 \emph{Green thick dotted line:} spiral galaxy at R=4 kpc;
 \emph{blue thick long-dashed line :} spiral galaxy at R=8 kpc;
 \emph{black thick short-dashed line:} spiral galaxy at R=14 kpc;
 \emph{red thick solid line:} spiral galaxy at R=18 kpc;
 \emph{thin dotted line:} magellanic irregular galaxy (LMC);
 \emph{thin dashed line:} elliptical galaxy. 
 \emph{Empty circles:} from various authors, see Vladilo 2002b and references therein;
 \emph{stars:} Prochaska et al. 2001b; 
 \emph{empty triangles:} Pettini et al. 1994;
 \emph{solid triangles:} Pettini et al. 1997;
 \emph{solid square:} Pettini et al. 2000;
 \emph{empty exagon:} Meyer et al. 1995;
 \emph{solid exagon:} P\'eroux et al. 2002;
 \emph{six-tips star:} Boiss\'e et al. 1998;
 \emph{asterisk:} Meyer \& York 1992.
 }	
\end{figure*}
%%%%%%%%%%%%%%%%%%%%%%%%%%%%%%%%%%%%%%%%%%%%%%%%%%%%%%%%%%%%%%%%%%%%%%%%%%%%%%%%%%%%%%%%%%%%%%%%%%%%%%%%%%%%

\subsection{The $\alpha$ vs Fe-peak ratio}

%%%%%%%%%%%%%%%%%%%%%%%%%%%%%%%%%%%%%%%%%%%%%%%%%%%%%%%%%%%%%%%%%%%%%%%%%%%%%%%%%%%%%%%%%%%%%%
Figure 4 shows the [Si/Fe] distribution versus redshift (above) and [Fe/H] (below),
as predicted by models of galaxies of
different morphological types and compared with uncorrected 
DLA data from various authors (Vladilo 2002b and references therein).
In particular, we plot predictions relative to a 
typical spiral like the MW at various galactocentric distances, 
to a typical elliptical, to a galaxy like the LMC and to a star-bursting dwarf which
has suffered only one burst of star formation.
We consider,
at the same time,  the [$\alpha$/Fe] ratios versus metallicity 
(usually Fe or Zn) distribution, 
which is completely independent of cosmological parameters.
As can be seen from the evolution as a function of redshift,
in all cases the predictions from the models show a short 
initial phase which is characterized by 
an overabundant [$\alpha$/Fe] ratio, during which type-II SNe are the 
main contributors to the chemical
enrichment of the ISM. This phase ends when type Ia SNe begin 
to appear.
In fact, after a given time, 
which is a strong function of the SN Ia 
progenitor lifetimes and of the star formation
history characterizing a given galactic morphological type
(see Matteucci \& Recchi, 2001),
type Ia SNe start to explode restoring the bulk of the
iron-peak elements, thus causing the decrease in the
[$\alpha$/Fe] ratios.
The disk-models at different radii experience the initial enhancements 
at various epochs: generally the 
smaller the radius at which the chemical evolution is computed, 
the earlier and the more pronounced 
is the peak in the [$\alpha$/Fe] ratio.
This is due to the different star formation histories occurring at different 
radii: a stronger star formation, 
owing to a higher gas density, is typical of the innermost regions, 
which mimic an elliptical-like star formation. On the other hand,
at larger radii the star formation is less intense, and this implies a
later and weaker peak in the [$\alpha$/Fe], similar to what happens in irregular galaxies.\\
It is also worth noticing that
the [$\alpha$/Fe] vs [Fe/H] plots 
show a quite different behaviour relative to the [$\alpha$/Fe] versus redshift plots.
In particular, while in the former plots the ellipticals show a higher 
[$\alpha$/Fe] ratio than the other galaxies over the complete [Fe/H] range, in the redshift plot the situation is reverted. This is due to the fact that the two kinds of plots give us different physical information; 
the [$\alpha$/Fe] versus [Fe/H] relation shows 
that the gas in a galaxy with an intense star formation should  have an
overabundance of $\alpha$-elements for a larger range of [Fe/H] values, owing to the fact that the large number of type II SNe
produces high [Fe/H] values in a time shorter than the typical timescale 
of type Ia SNe. 
In other words, a short redshift interval can contain a large interval of [Fe/H] values. Therefore, the plots as functions 
of redshift (cosmic time) and [Fe/H] behave in the opposite way.\\
In the case of the [Si/Fe] ratio we can rely on a large data sample. 
Usually both Si and Fe abundances can easily be assessed in DLA systems, 
none of the two being affected by
Lyman-$\alpha$ contamination or severe saturation. 
The only concern is dust depletion, which 
usually affects both elements, although in different amounts.
In figure 4 we show two plots: [Si/Fe] vs z (upper panel) and [Si/Fe] vs [Fe/H] (lower panel).
The data shown in the figures are not yet dust corrected.
In both plots the majority of observed systems 
shows strong signs of Si enhancement. 
In the [Si/Fe]-redshift evolution no morphological type can 
satisfactorily reproduce the data.
If we consider the [Si/Fe] vs [Fe/H] diagram, all but one of the 
most metal-poor systems 
(the ones with [Fe/H]$<$-1.8) are fairly consistent with spiral/irregular 
patterns.
For [Fe/H]$>$-1.8, we have several values reproduced by the elliptical 
model, a small fraction of points
in-between the elliptical and the spiral/irregular curves and another 
fraction of data reproduced by 
spirals and irregulars.
However this observed trend would reflect a real overabundance of Si with respect 
to the 
iron-peak elements only in the complete absence of dust.
If we consider that the dust content scales with the metallicity, 
the only systems where the influence of dust depletion can be small 
are the most metal-poor,
as noticed by Pettini et al. (1995).

%%%%%%%%%%%%%%%%%%%%%%%%%%%%%%%%%%%%%%%%%%%%%%%%%%%%%%%%%%%%%%%%%%%%%%%%%%%%%%%%%%%%%%%%%%%%%%%%%%%%%%%%%%%%%
\begin{figure*}
\centering
\vspace{0cm}
\psfig{file=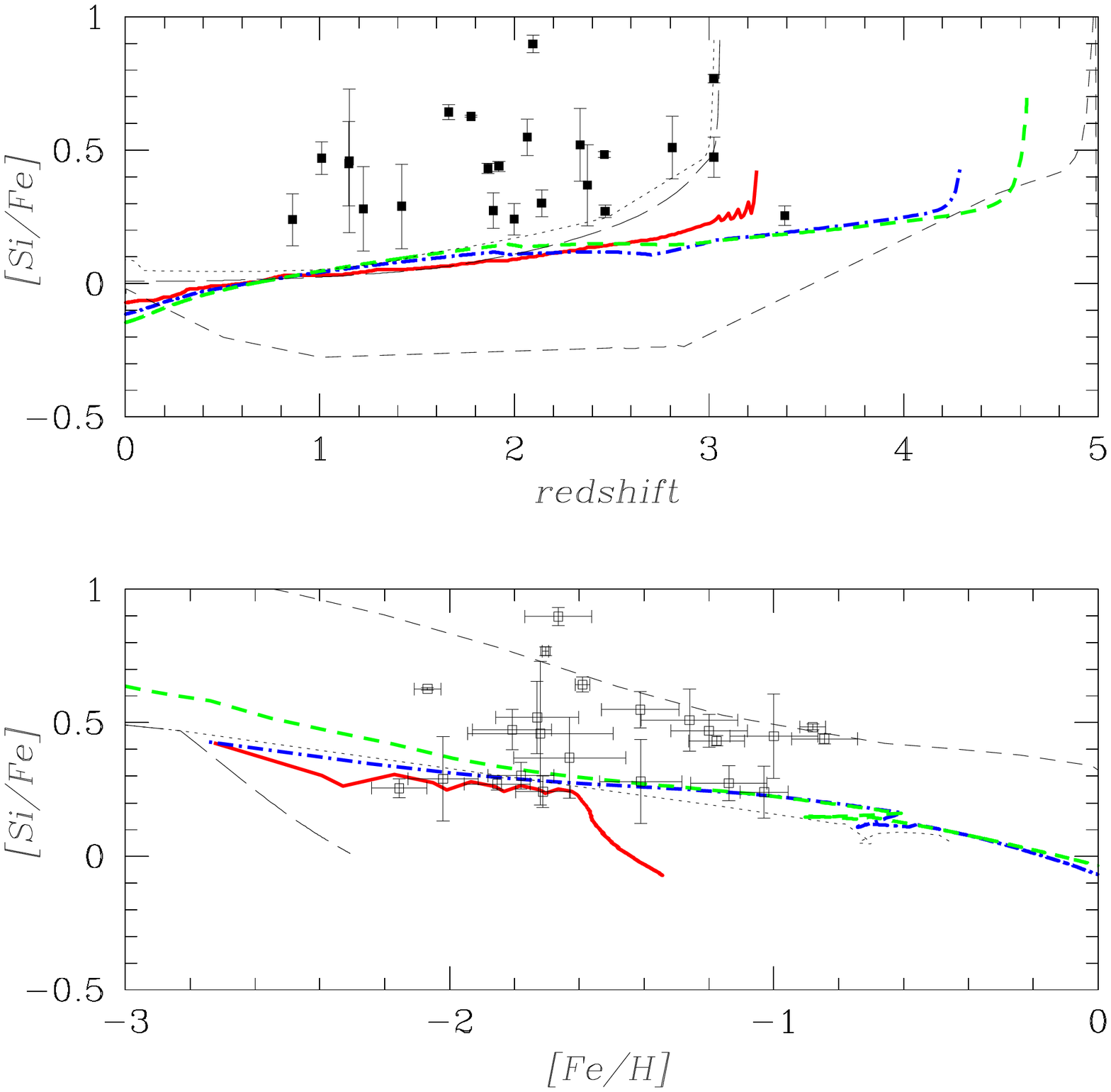,height=10cm,width=10cm}
\caption[]{Comparisons between theoretical predictions and observations for [Si/Fe] vs z (above) and vs [Fe/H]
(below) in case of {\bf not dust-corrected} data. The redshift of galaxy formation is $z_{f}=5$.
 \emph{Green thick dashed line:} spiral at R=4 kpc;
 \emph{blue thick dot-dashed line:} spiral at R=8 kpc;
 \emph{red thick solid line:} spiral at R=18 kpc;
 \emph{thin dotted line:} irregular;
 \emph{thin short-dashed line:} elliptical;
 \emph{thin long-dashed line:} model for a starburst galaxy with a single 
burst of star 
formation at age of 1 Gyr and with duration 
$\Delta t_{burst}=0.05$ Gyr.
  Data are taken from various authors (see Vladilo 2002b and references therein).
 
 }	
\end{figure*}
%%%%%%%%%%%%%%%%%%%%%%%%%%%%%%%%%%%%%%%%%%%%%%%%%%%%%%%%%%%%%%%%%%%%%%%%%%%%%%%%%%%%%%%%%%%%%%%%%%%%%%%%%%%%
%%%%%%%%%%%%%%%%%%%%%%%%%%%%%%%%%%%%%%%%%%%%%%%%%%%%%%%%%%%%%%%%%%%%%%%%%%%%%%%%%%%%%%%%%%%%%%%%%%%%%%%%%%%%%
\begin{figure*}
\centering
\vspace{0cm}
\psfig{file=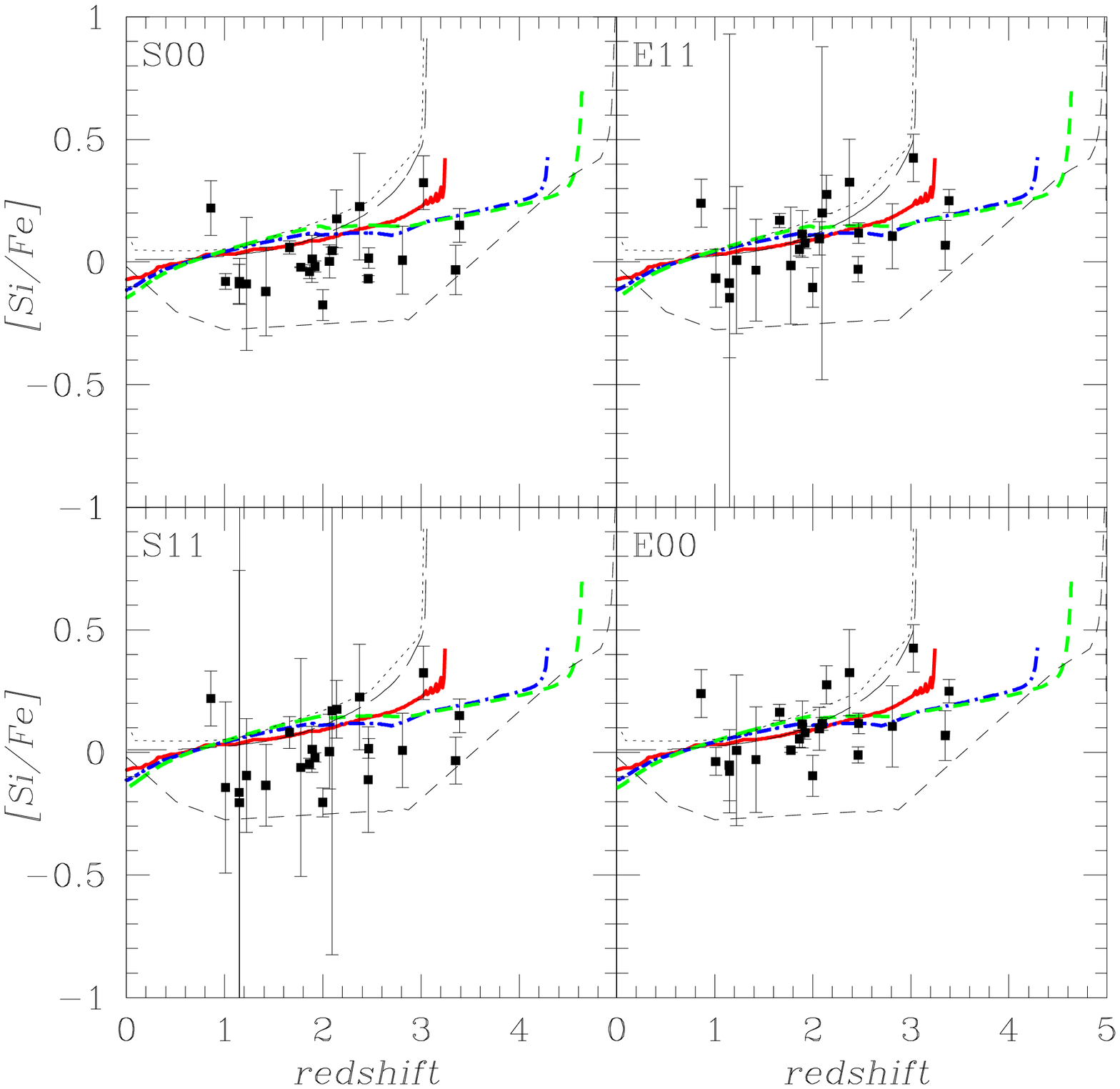,height=10cm,width=10cm}
\caption[]{Comparisons between theoretical predictions and observations for [Si/Fe] vs z in case of {\bf dust-corrected} data. The redshift of galaxy formation is $z_{f}=5$.
Four different assumptions have been made on the parameters describing dust. 
The curves have the same meaning as in Figure 4.
%  \emph{Thick dashed line:} spiral at R=4 kpc;
% \emph{thick dot-dashed line:} spiral at R=8 kpc;
% \emph{thick solid line:} spiral at R=18 kpc;
% \emph{thin dotted line:} irregular;
% \emph{thin short-dashed line:} elliptical;
% \emph{thin long-dashed line:} starburst.
Data taken from various authors ( see Vladilo 2002b and references therein).
  }	
\end{figure*}
%%%%%%%%%%%%%%%%%%%%%%%%%%%%%%%%%%%%%%%%%%%%%%%%%%%%%%%%%%%%%%%%%%%%%%%%%%%%%%%%%%%%%%%%%%%%%%%%%%%%%%%%%%%%%%%%%%%%%%%%%%%%%%%%%%%%
\begin{figure*}
\centering
\vspace{0cm}
\psfig{file=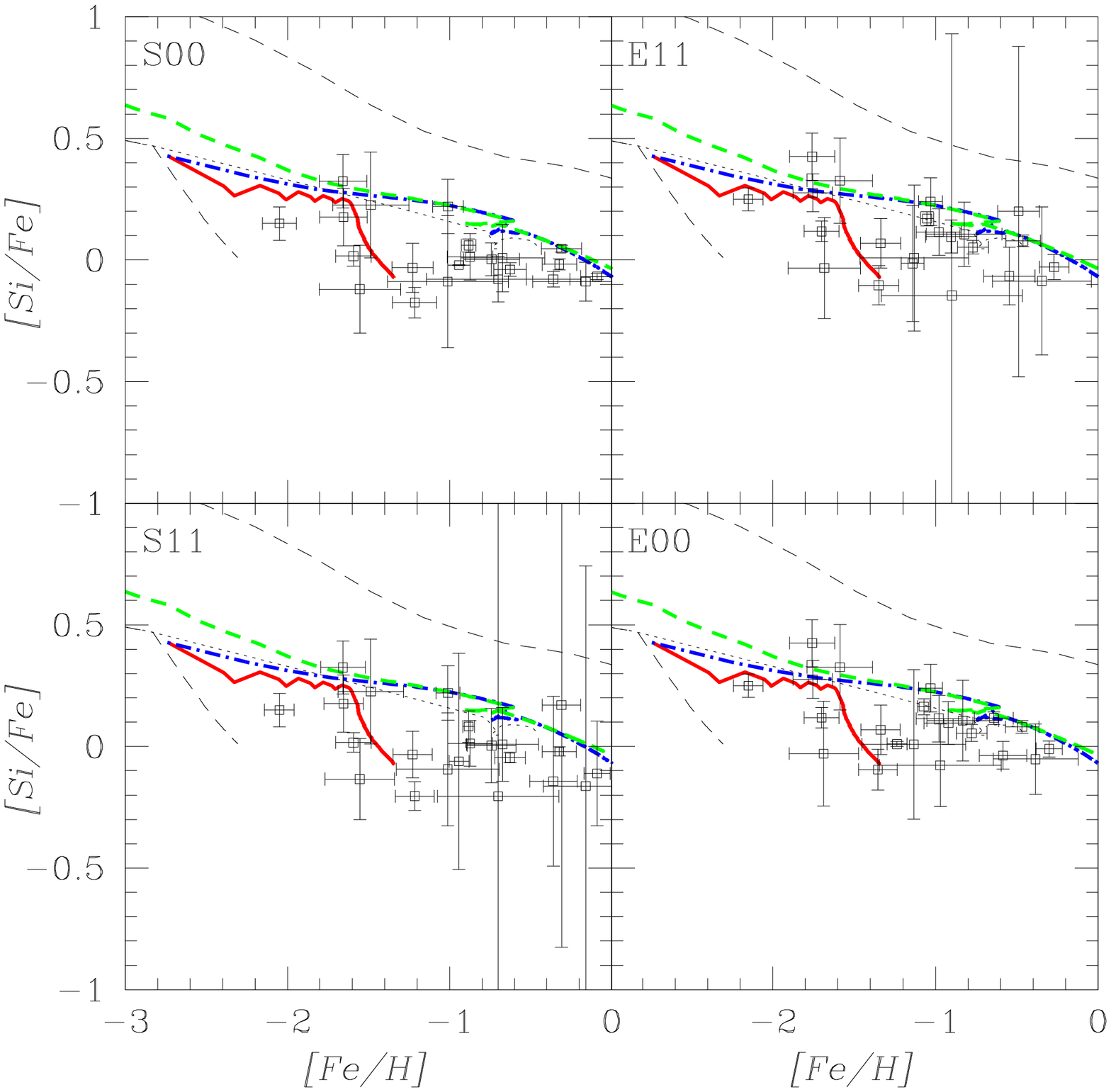,height=10cm,width=10cm}
\caption[]{Comparisons between theoretical predictions and observations for [Si/Fe] vs [Fe/H] in case of {\bf dust-corrected} data. 
Four different assumptions have been made on the parameters describing dust. The curves have the same meaning as in Figure 4.
% \emph{Thick dashed line:} spiral at R=4 kpc;
% \emph{thick dot-dashed line:} spiral at R=8 kpc;
% \emph{thick solid line:} spiral at R=18 kpc;
% \emph{thin dotted line:} irregular;
% \emph{thin short-dashed line:} elliptical;
% \emph{thin long-dashed line:} starburst.
Data taken from various authors ( see Vladilo 2002b and references therein).
 }	
\end{figure*}
%%%%%%%%%%%%%%%%%%%%%%%%%%%%%%%%%%%%%%%%%%%%%%%%%%%%%%%%%%%%%%%%%%%%%%%%%%%%%%%%%%%%%%%%%%%%%%%%%%%%%%%%%%%%%%%%%%%%%%%%%%%%%%%%%%%%
%%%%%%%%%%%%%%%%%%%%%%%%%%%%%%%%%%%%%%%%%%%%%%%%%%%%%%%%%%%%%%%%%%%%%
%\begin{table*}
%\vspace{0cm}
%\begin{flushleft}
%\caption[]{Assumed parameters to compute dust corrections according to the method described in Vladilo 2002b.}
%\begin{tabular}{l|lll}
%\noalign{\smallskip}
%\hline
%\hline
%\noalign{\smallskip}
%Model & [Zn/Fe]  & $\epsilon_{Zn}$ & $\epsilon_{Si}$\\
%\noalign{\smallskip}
%\hline
%\noalign{\smallskip}
% S00 & 0 & 0 & 0\\
% S11 & 0 & 1 & 1\\
% E00 & 0.1 & 0 & 0\\
% E11 & 0.1 & 1 & 1\\
%\noalign{\smallskip}
%\hline
%\noalign{\smallskip}
%\hline
%\end{tabular}
%\end{flushleft}
%\end{table*}
%%%%%%%%%%%%%%%%%%%%%%%%%%%%%%%%%%%%%%%%%%%%%%%%%%%%%%%%%%%%%%%%%%%%%%%%%%%%%%%%
Figures 5 and 6 show the results of the comparison between the model 
predictions and observations for the 
[Si/Fe] vs z  and [Fe/H], respectively, 
in the case of {\bf dust-corrected} data. The corrections have 
been calculated 
according to the dust model discussed in Vladilo (2002a,b), where
a scaling law for interstellar depletion has been derived allowing
the chemical composition of dust to vary according to changes both 
in the dust-to-metals ratio and/or in the abundances in the medium. 
Here we present four different assumptions about two parameters 
connected to the dust; 
one parameter is the [Zn/Fe] ratio and the other is $\epsilon$, 
as defined by Vladilo, which indicates the percent variation of 
the abundance of a given element in the dust which occurs in correspondance 
of a percent variation of the abundance of the same element in the medium
(see eq. 14 of Vladilo 2002a for the definition of this parameter).
We refer to the four cases as model S00, S11, E00
and E11. Letters S and E refer to the adopted assumptions 
for [Zn/Fe] (S$\rightarrow [Zn/Fe]=0.0$, 
E$\rightarrow [Zn/Fe]=0.1$) The first number refers to the adopted 
$\epsilon$ value for the element Zn and the second number is the $\epsilon$
value for the element Si.
In all these cases the $\alpha$-enhancement is  considerably less 
pronounced than previously
or even absent.
Generally, the dust corrections reduce the
[$\alpha$/Fe] ratios by
$\sim 0.3-0.5$ dex, 
which is enough to lower them to the solar value.\\
The predicted [Si/Fe] versus redshift and [Fe/H] relations in Figures 5 and 6
show a much better agreement with the data than in Figure 4. 
In particular, most of the data now seem to be well reproduced by
models of spirals and
irregulars, whereas elliptical galaxies can never fit the points. 
This is because in the very early 
phases the ellipticals reach too high values of the [Si/Fe] 
ratio. Later on, the release of the 
Fe-peak elements by type Ia SN 
is strong enough to lower [Si/Fe] to values strongly undersolar, 
too small if compared with the other
galaxies.
The subsequent increase of [Si/Fe] at very late times is due to the fact 
that low mass stars ($\sim 1 M_{\odot}$) are dying and restoring unprocessed material
which contains the [Si/Fe] present in the ISM when they formed, and
therefore enriched in Si relative to Fe.
Spiral galaxies evolve
with smoother and less intense star formation histories than ellipticals 
and never reach such 
high enhancements of the [$\alpha$/Fe] ratios, 
but they can have high final metallicities, 
especially in the most internal regions,
thus they can reproduce the observed trend with noticeable accuracy at 
all metallicities.
In principle it is very difficult to distinguish between typical low, 
smooth star formation patterns in
irregular galaxies and in spiral disks considering only the pure redshift 
evolution of the abundance ratios such as
[Si/Fe] since, as can be seen from the two figures, an irregular galaxy 
with a huge gas content and a very low
star formation can show an evolution of [Si/Fe] which is essentially 
identical to that of the external region of a spiral  galaxy.
Furthermore, the chemical evolution histories of a spiral galaxy 
computed at different radii can be very similar
to each others, with the only difference represented by the height 
and the epoch of the initial peak of
$\alpha$-enhancement, which is usually very short (as can be seen in figure 4, since for $z<2$ all the curves of
the spiral and of the irregular are overlapping).
On the other hand, differences in the evolution patterns are more evident in the plots as functions of [Fe/H].
This is a clear example of the power of the  
the [$\alpha$/Fe]-[Fe/H] diagrams 
in distinguishing between the different chemical evolution 
histories and in particular to understand which
morphological types best represent the DLA population.\\
%%%%%%%%%%%%%%%%%%%%%%%%%%%%%%%%%%%%%%%%%%%%%%%%%%%%%%%%%%%%%%%%%%%%%%%%%%%%%%%%%%%%%%%%%%%%%%%%%%%%%%%%%%%%%
\begin{figure*}
\centering
\vspace{0cm}
\psfig{file=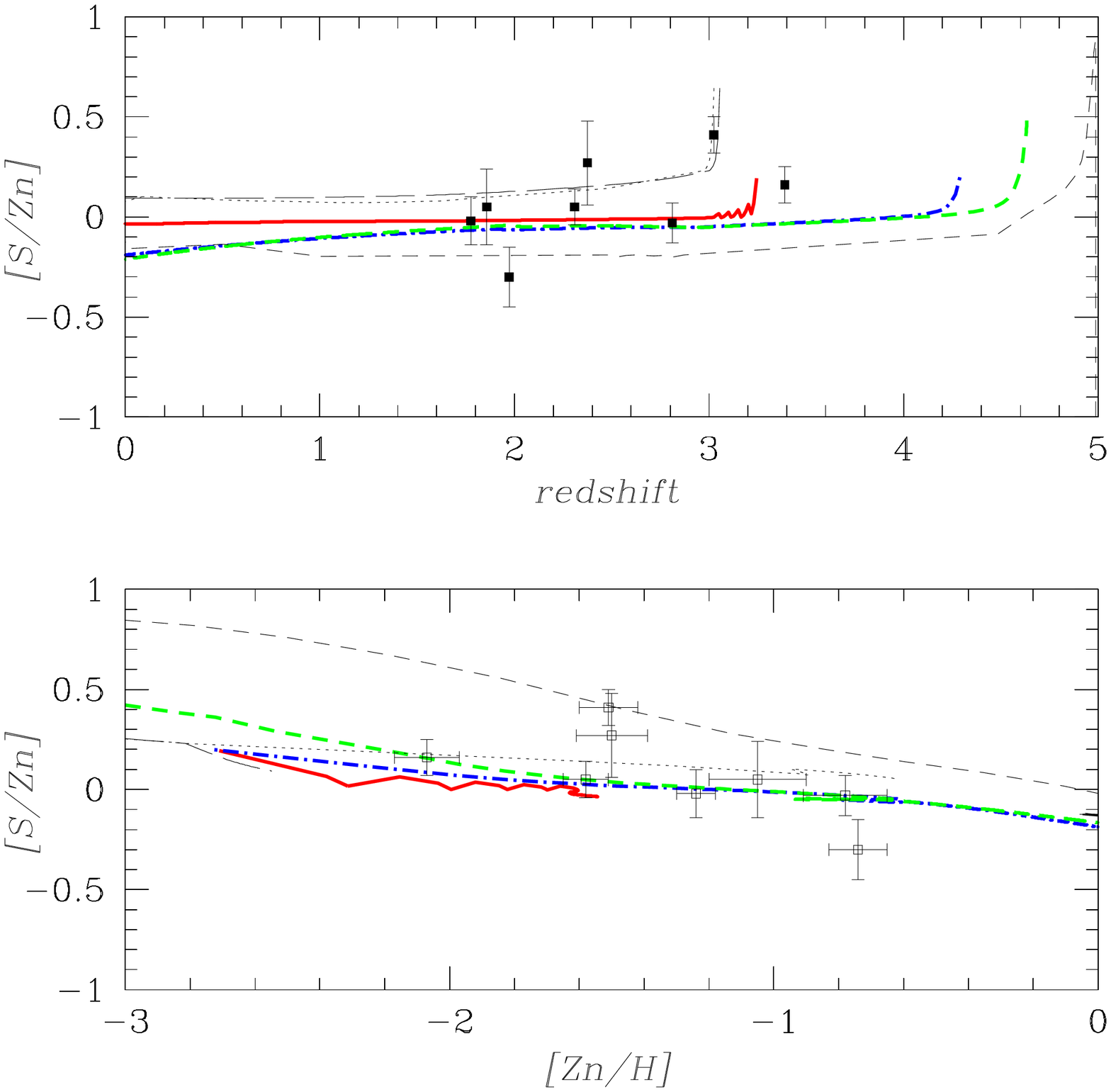,height=10cm,width=10cm}
\caption[]{Comparisons between theoretical predictions and observations for [S/Zn] vs z (upper panel) and vs [Zn/H]
(bottom panel). The reshift of galaxy formation is $z_{f}=5$.The curves have same meaning as in Figure 4. 
% \emph{Thick dashed line:} spiral at R=4 kpc;
% \emph{thick dot-dashed line:} spiral at R=8 kpc (SN);
% \emph{thick solid line:} spiral at R=18 kpc;
% \emph{thin dotted line:} irregular;
% \emph{thin short-dashed line:} elliptical;
% \emph{thin long-dashed line:} starburst.
Compilation of data taken from Centuri\'on et al. (2002).
 
 }	
\end{figure*}
%%%%%%%%%%%%%%%%%%%%%%%%%%%%%%%%%%%%%%%%%%%%%%%%%%%%%%%%%%%%%%%%%%%%%%%%%%%%%%%%%%%%%%%%%%%%%%%%%%%%%%%%%%%%
Fig. 7 shows the [S/Zn] distribution versus redshift (above) and versus [Zn/H]
(below) for different morphological types compared with measurements in DLAs. 
Notwithstanding the large dispersion in the redshift distribution, a general trend is clearly
observable, in which most of the points are fitted by spiral/irregular curves. 
Once again, the evolution of the [S/Zn]
ratio for elliptical galaxies is too sharp to reproduce the bulk of the data, 
since it is too high in the initial
phase and falls to too low values already at $z<4$.\\
Again, the [S/Zn] vs [Zn/H] plot is a more reliable 
chemical evolution diagnostic than the [S/Zn] vs. redshift evolution.
We note that both irregulars and spirals (at inner radii and in 
a region equivalent  to the solar neighbourhood)
can reproduce satisfactorily most of the observed values.
Only one point, corresponding to the most enhanced value, is consistent 
solely with the elliptical evolution, whereas another DLA system has a
[S/Zn] value lower than the predictions of all the models considered here.
We stress the fact that in general the results concerning 
the [S/Zn] vs z, [Zn/H]
distributions tend to confirm what has been concluded in the case of dust-corrected [Si/Fe] vs z, [Fe/H], as expected, since both Zn and S should not be 
affected by dust depletion.

%%%%%%%%%%%%%%%%%%%%%%%%%%%%%%%%%%%%%%%%%%%%%%%%%%%%%%%%%%%%%%%%%%%%%%%%%%%%%%%%%%%%%%%%%%%%%%%%%%%%%%%%%%%%%
\begin{figure*}
\centering
\vspace{0cm}
\psfig{file=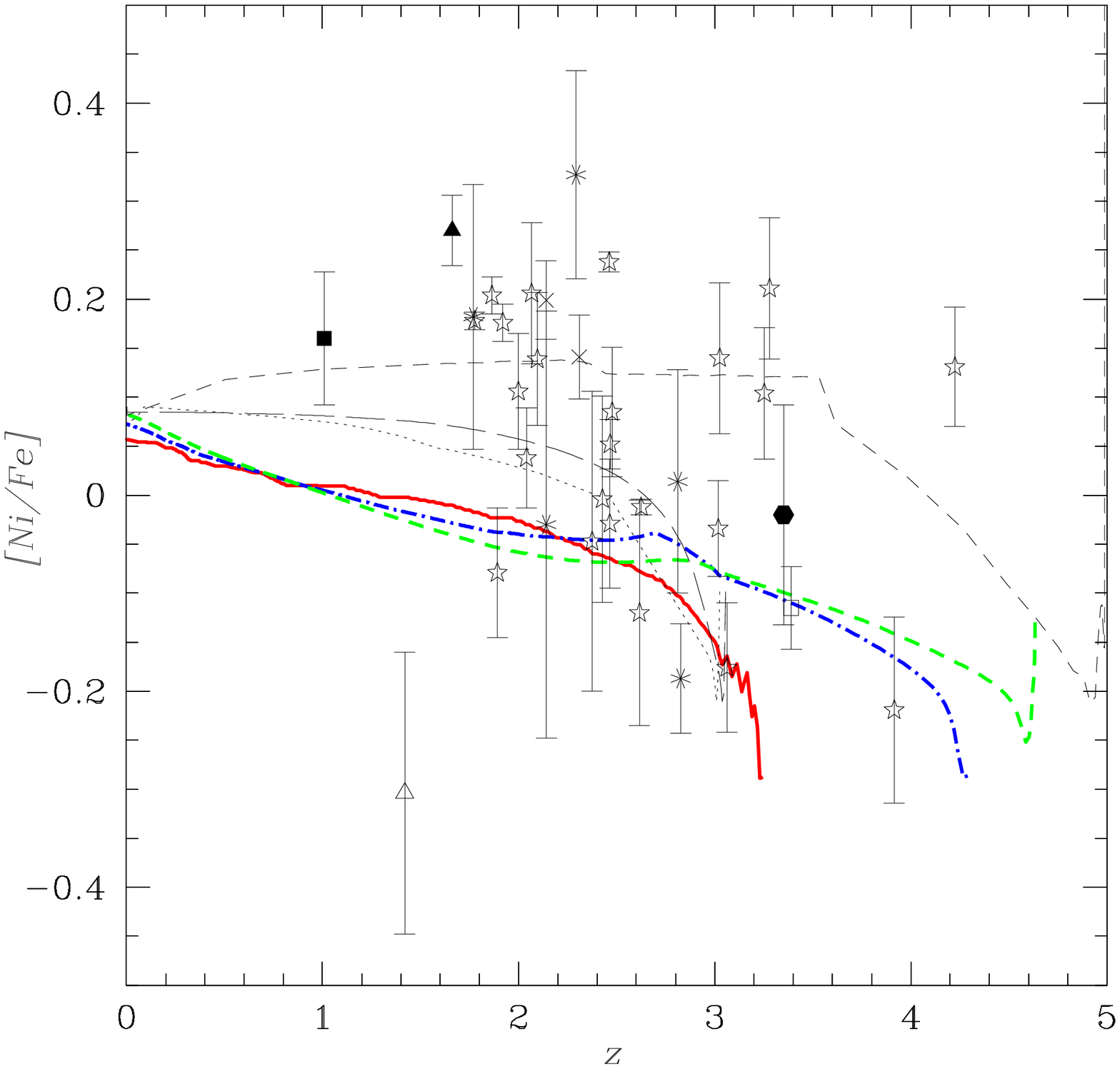 ,height=10cm,width=10cm}
\caption[]{Comparisons between theoretical predictions and observations 
for [Ni/Fe] vs redshift in 
case of galaxy formation at $z_{f}=5$. The meaning of the curves is 
the same as in Figure 4.
 \emph{Open square:} Molaro et al. 2000;
 \emph{crosses:} Prochaska \& Wolfe 1999;
 \emph{stars:} Prochaska et al. 2001b;
 \emph{asterisks:} Lu et al. 1996;
 \emph{solid square:} Pettini et al. 2000;
 \emph{solid triangle:} L\'opez et al. 1999;
 \emph{solid exagon:} P\'eroux et al. 2002;
 \emph{open triangle:} Pettini et al. 1999.
}

\end{figure*}
%%%%%%%%%%%%%%%%%%%%%%%%%%%%%%%%%%%%%%%%%%%%%%%%%%%%%%%%%%%%%%%%%%%%%%%%%%%%%%%%%%%%%%%%%%%%%%%%%%%%%%%%%%%%
%%%%%%%%%%%%%%%%%%%%%%%%%%%%%%%%%%%%%%%%%%%%%%%%%%%%%%%%%%%%%%%%%%%%%%%%%%%%%%%%%%%%%%%%%%%%%%%%%%%%%%%%%%%%%
\begin{figure*}
\centering
\vspace{0cm}
\psfig{file=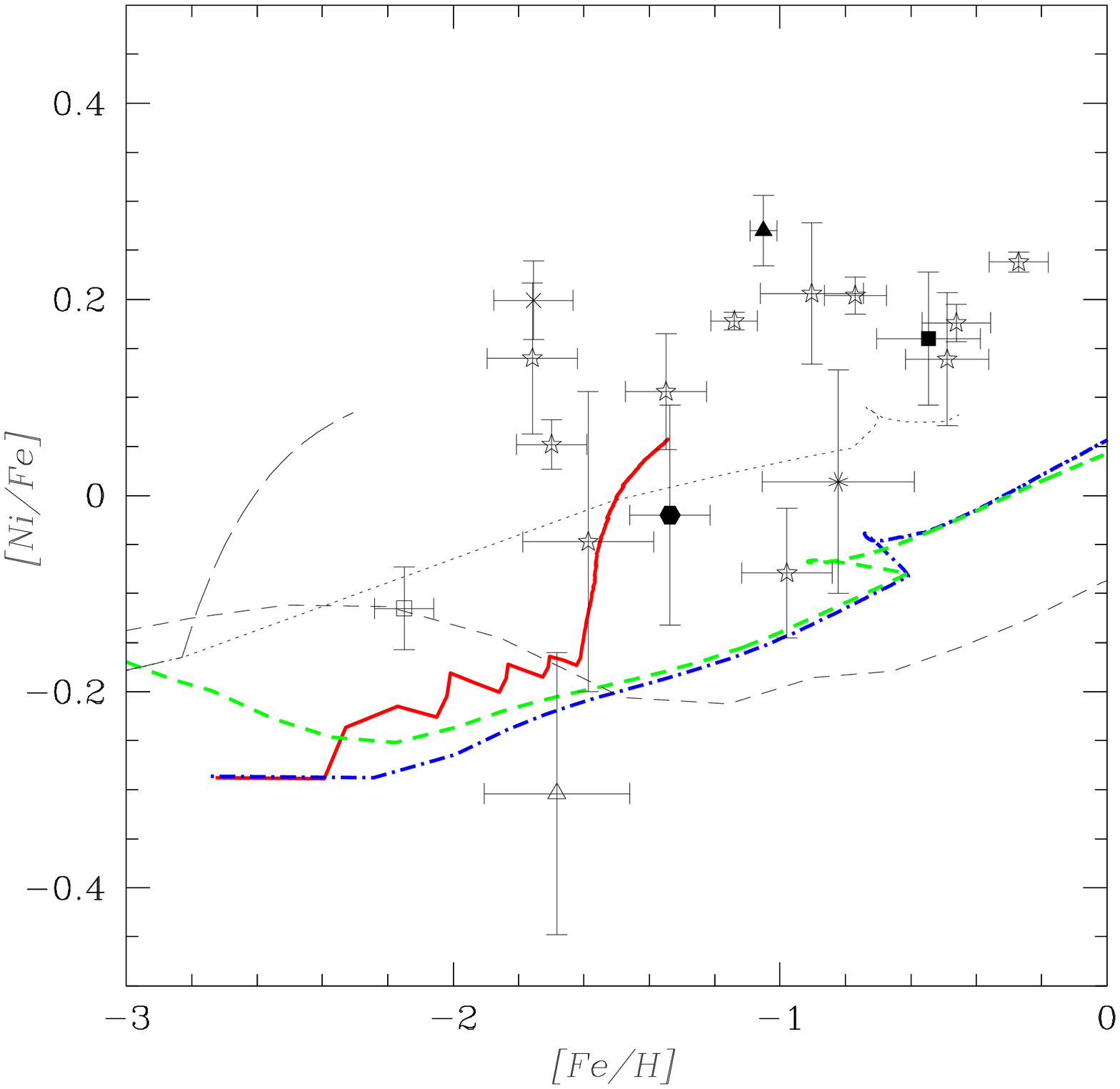 ,height=10cm,width=10cm}
\caption[]{Comparisons between theoretical predictions and observations 
for [Ni/Fe] vs dust-corrected [Fe/H]. The meaning of the curves is 
the same as in Figure 4.
Symbols are as in figure 8.

}

\end{figure*}
%%%%%%%%%%%%%%%%%%%%%%%%%%%%%%%%%%%%%%%%%%%%%%%%%%%%%%%%%%%%%%%%%%%%%%%%%%%%%%%%%%%%%%%%%%%%%%%%%%%%%%%%%%%%
%%%%%%%%%%%%%%%%%%%%%%%%%%%%%%%%%%%%%%%%%%%%%%%%%%%%%%%%%%%%%%%%%%%%%%%%%%%%%%%%%%%%%%%%%%%%%%%%%%%%%%%%%%%%%
\begin{figure*}
\centering
\vspace{0cm}
\psfig{file=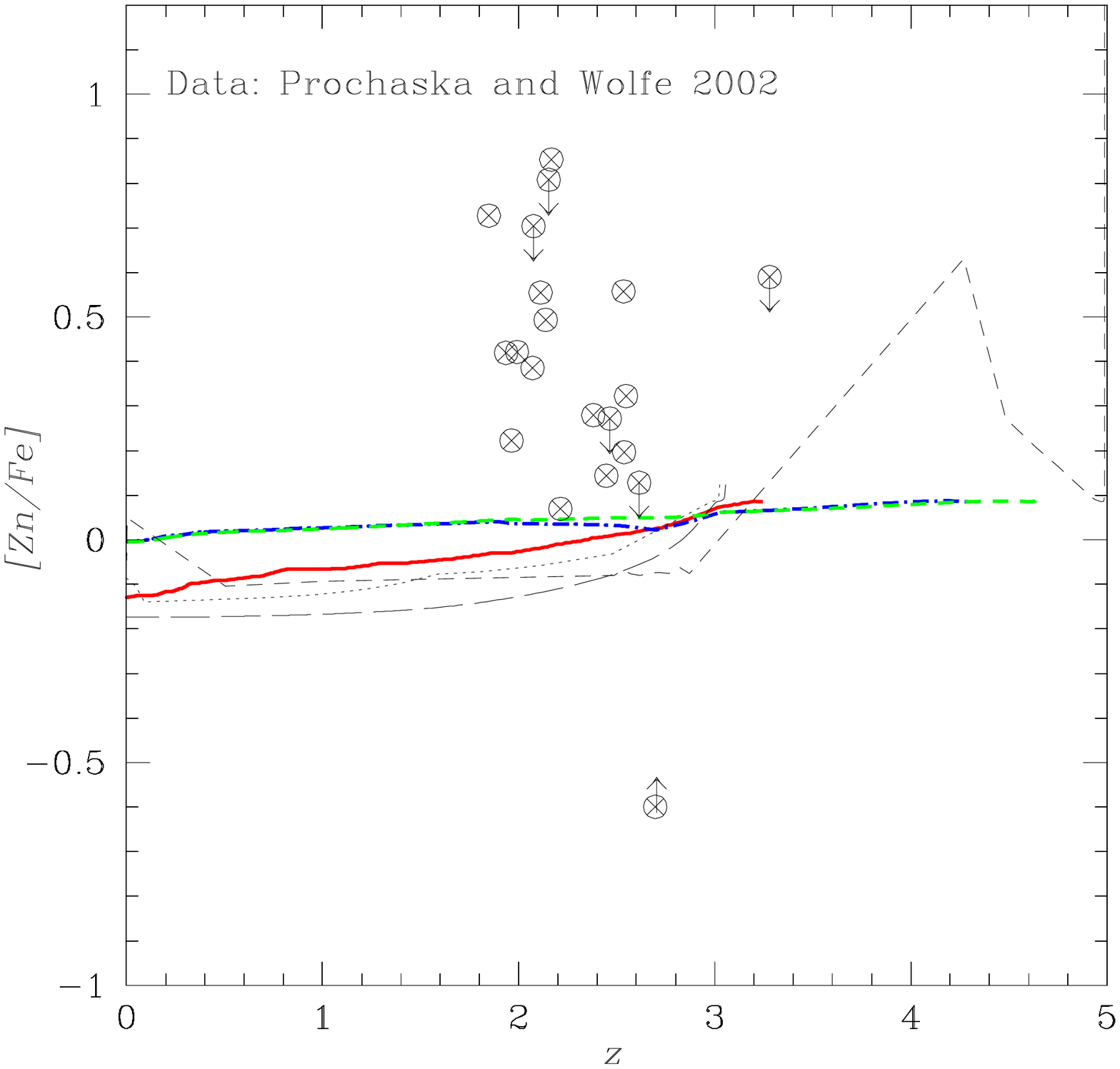 ,height=10cm,width=10cm}
\caption[]{Comparisons between theoretical predictions and observations for [Zn/Fe] 
vs redshift in 
case of galaxy formation at $z_{f}=5$. The meaning of the curves is the same as in 
Figure 4.
% \emph{Thick dashed line:} spiral at R=4 kpc;
% \emph{thick dot-dashed line:} spiral at R=8 kpc (SN);
% \emph{thick solid line:} spiral at R=18 kpc;
% \emph{thin dotted line:} irregular;
% \emph{thin short-dashed line:} elliptical;
% \emph{thin long-dashed line:} starburst.
}	
\end{figure*}
%%%%%%%%%%%%%%%%%%%%%%%%%%%%%%%%%%%%%%%%%%%%%%%%%%%%%%%%%%%%%%%%%%%%%%%%%%%%%%%%%%%%%%%%%%%%%%%%%%%%%%%%%%%%
%%%%%%%%%%%%%%%%%%%%%%%%%%%%%%%%%%%%%%%%%%%%%%%%%%%%%%%%%%%%%%%%%%%%%%%%%%%%%%%%%%%%%%%%%%%%%%%%%%%%%%%%%%%%%
\begin{figure*}
\centering
\vspace{0cm}
\psfig{file=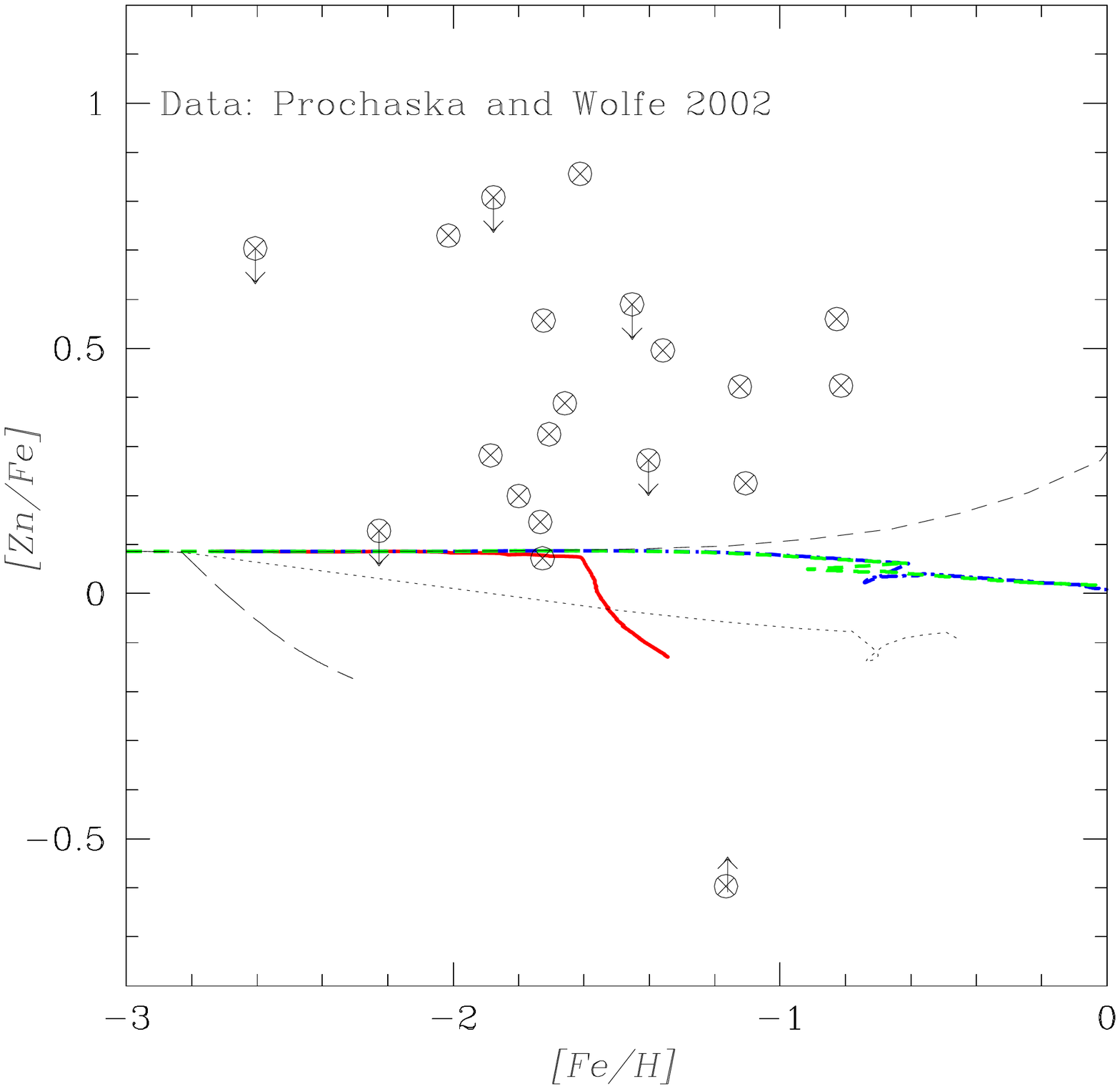 ,height=10cm,width=10cm}
\caption[]{Comparisons between theoretical predictions and observations for [Zn/Fe] vs [Fe/H]. The meaning of the curves is the same as in Figure 4.
% \emph{Thick dashed line:} spiral at R=4 kpc;
% \emph{thick dot-dashed line:} spiral at R=8 kpc (SN);
% \emph{thick solid line:} spiral at R=18 kpc;
% \emph{thin dotted line:} irregular;
% \emph{thin short-dashed line:} elliptical;
% \emph{thin long-dashed line:} starburst.
}	
\end{figure*}
%%%%%%%%%%%%%%%%%%%%%%%%%%%%%%%%%%%%%%%%%%%%%%%%%%%%%%%%%%%%%%%%%%%%%%%%%%%%%%%%%%%%%%%%%%%%%%%%%%%%%%%%%%%%

\subsection{The Fe-peak elements}
In this case we study the evolution with redshift and metallicity of 
[Ni/Fe] and [Zn/Fe] ratios.
Figures 8-11 show the results. 
In spite of the scatter of the points around the solar value, 
the majority of them appears moderately enhanced in both cases, 
at variance with what can be observed in the Galactic ISM.\\ 
It is worth noting that,
when the most recent values of Ni oscillator strengths are taken into
account,
the parameters for dust correction of Ni and Fe happen to be
almost identical (Vladilo 2002b; Table 1) and the Ni/Fe ratios
are essentially free of dust effects.
Therefore,
the application of the dust correction method does not change the values of
the ratios.
In Figure 8 we plot the available [Ni/Fe] measurements without dust corrections.
In Fig. 9 we plot the measurements of
[Ni/Fe] versus [Fe/H], where only the [Fe/H] values have been
corrected for dust depletion
using the set of parameters E11.
Very similar results are obtained using the sets of parameters
S00, S11 and E00.
The sample of Fig. 9 is smaller than that of Fig. 8
because it includes only the DLAs for which
the dust correction procedure can be applied.
In the redshift evolution plot of the [Ni/Fe] (Fig. 8), 
the lowest values are well reproduced by the 
irregular-spiral curves, but there are some points 
which are
consistent with an elliptical-like evolution; the highest values instead
are off-scale 
for any galactic model. However, this 
behaviour might be due to the uncertainties present in 
the theoretical yields of 
Fe-peak elements
(Matteucci et al. 1993; Timmes et al. 1995; Umeda \& Nomoto, 2002).
\\
In the [Zn/Fe] case the observed enhancement is even stronger, and 
most of the 
data would be consistent with the elliptical peak, if it occurred between 
redshifts 2 and 3, but these data are not dust corrected and therefore 
no firm conclusions can be drawn.
The nearly flat behaviour of the [Zn/Fe] ratio predicted for the irregular, 
inner disk, solar neighbourhood and elliptical models is essentially
due to the fact that, according to the nucleosynthesis 
prescriptions assumed here, 
Zn and Fe follow each-other since 
both are mainly produced in SNe Ia. 
The outer regions of the disk have a different behaviour because 
of the assumed 
threshold density for the star formation rate.
This effect is more pronounced in the external regions 
than in the inner regions, 
because of the gas density profile, 
which is strongly decreasing 
with galactocentric radius and hence closer to the threshold value in the outermost regions of the disk.
As a consequence of this, the star formation regime is a starburst-like.
The existence of uncertainties in the Ni nucleosynthesis prescriptions
prevents us from drawing firm conclusions about the behaviour of the [Ni/Fe]
vs. [Fe/H] and redshift. On the other hand, the predictions of our models
indicate that, at least to first order, the productions of Zn and Fe 
can be considered strictly linked together and independent from metallicity.
Based on this result, the observed spread in the [Zn/Fe] 
values may be ascribed more to effects of dust 
(such as differential dust depletion of Fe and Zn, differential dust 
pollution along different lines-of-sight, different amounts of dust
within DLA systems) than to pure nucleosynthesis.
%%%%%%%%%%%%%%%%%%%%%%%%%%%%%%%%%%%%%%%%%%%%%%%%%%%%%%%%%%%%%%%%%%%%%%%%%%%%%%%%%%%%%%%%%%%%%%%%%%%%%%%%%%%%%
\begin{figure*}
\centering
\vspace{0cm}
\psfig{file=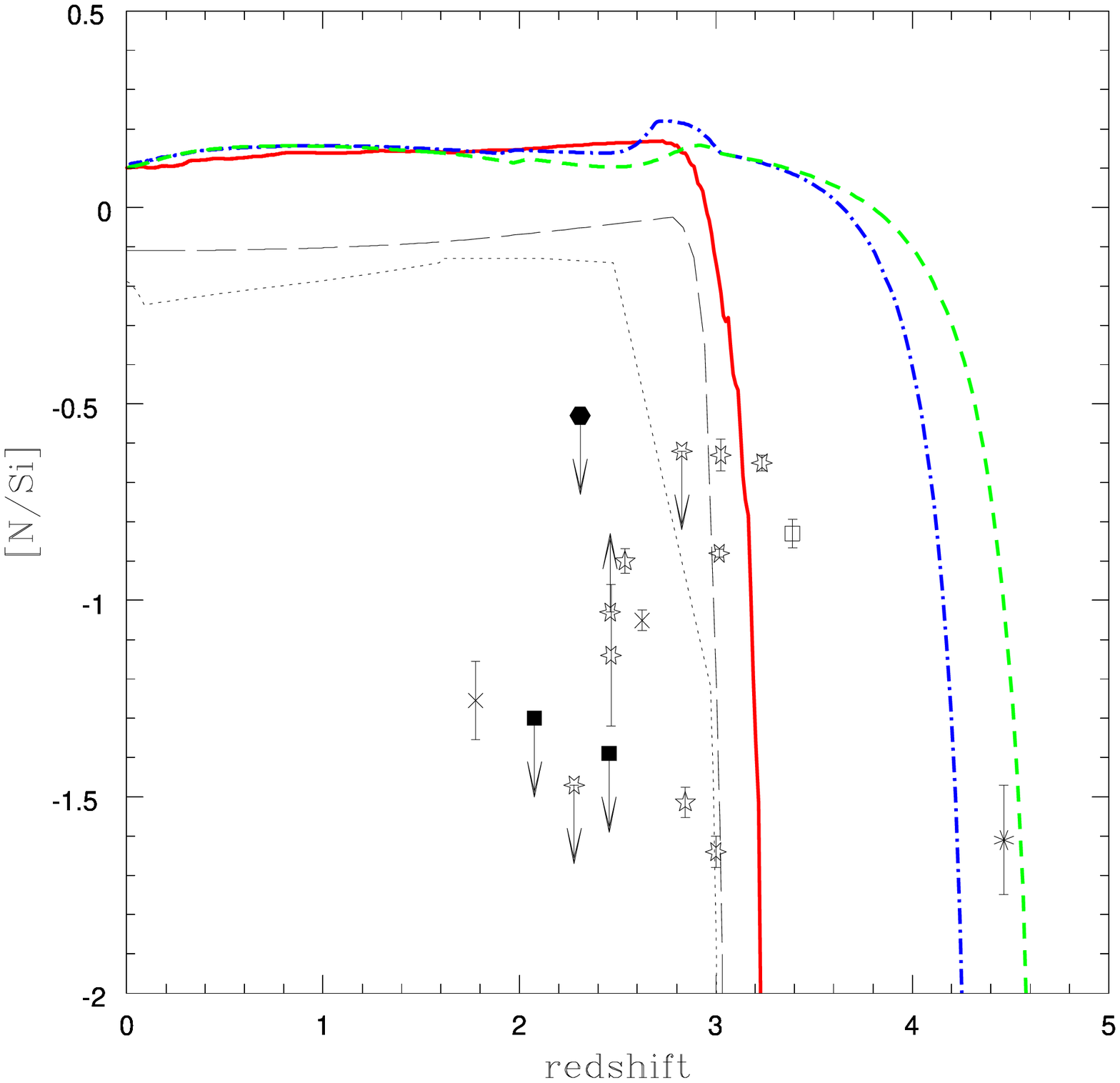,height=10cm,width=10cm}
\caption[]{Comparisons between theoretical predictions and observations 
for [N/Si] vs redshift in 
case of galaxy formation at $z_{f}=5$.
 \emph{Green thick dashed line:} spiral galaxy at R=4 kpc;
 \emph{blue thick dot-dashed line:} spiral galaxy at R=8 kpc;
 \emph{red thick solid line:} spiral galaxy at R=18 kpc;
 \emph{thin dotted line:} magellanic irregular galaxy (LMC);
% \emph{thin short-dashed line:} elliptical;
 \emph{thin long-dashed line:} starburst galaxy as described in Figure 4.
The observational data are:
 \emph{Empty squares:} Molaro et al. (2000);
 \emph{crosses:} Prochaska \& Wolfe (1999);
 \emph{asterisks:} Dessauges-Zavadsky et al. (2001);
% \emph{four-points stars:} Levshakov et al. (2002);
 \emph{five-points stars:} Prochaska et al. (2001b);
 \emph{six-points stars:} Prochaska et al. (2002b);
 \emph{exagons:} Centuri\'on et al. 1998;
 \emph{solid squares:} Pettini et al. 2002.
}

\end{figure*}
%%%%%%%%%%%%%%%%%%%%%%%%%%%%%%%%%%%%%%%%%%%%%%%%%%%%%%%%%%%%%%%%%%%%%%%%%%%%%%%%%%%%%%%%%%%%%%%%%%%%%%%%%%%%

%%%%%%%%%%%%%%%%%%%%%%%%%%%%%%%%%%%%%%%%%%%%%%%%%%%%%%%%%%%%%%%%%%%%%%%%%%%%%%%%%%%%%%%%%%%%%%%%%%%%%%%%%%%%%
\begin{figure*}
\centering
\vspace{0cm}
\psfig{file=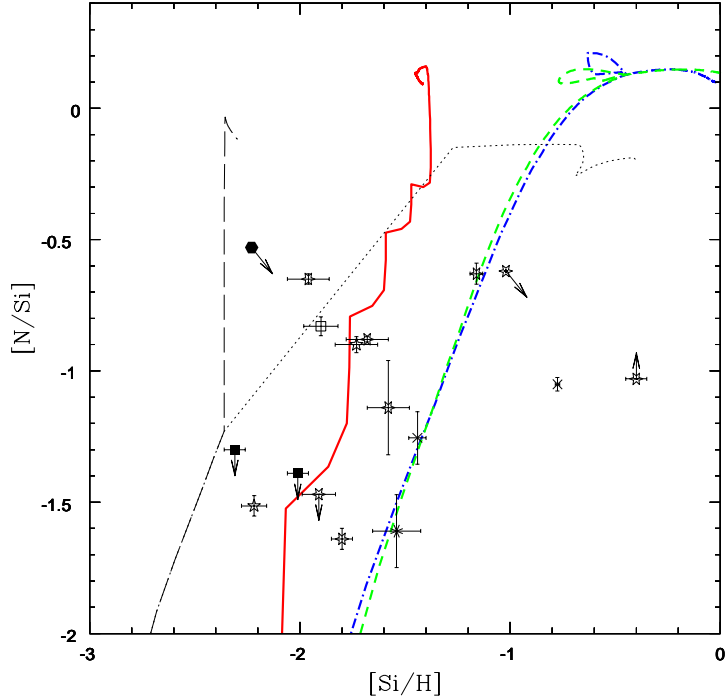,height=10cm,width=10cm}
\caption[]{Comparisons between theoretical predictions and observations 
for [N/Si] vs [Si/H]. The meaning of the curves as well as the symbols are
as in the previous figure.
}

\end{figure*}
%%%%%%%%%%%%%%%%%%%%%%%%%%%%%%%%%%%%%%%%%%%%%%%%%%%%%%%%%%%%%%%%%%%%%%%%%%%%%%%%%%%%%%%%%%%%%%%%%%%%%%%%%%%%

\subsection{The [N/$\alpha]$ ratio}
%The [N/$\alpha$] ratio can provide very useful hints on galactic 
%chemical evolution because of the different processes producing 
%nitrogen and $\alpha$-elements.
%N is usually 
%produced through the CNO cycle in stars of any mass according 
%to their initial metal content (in particular C and O), 
%so that the amount of N 
%sinthesized strongly depends on the initial metallicity of the star. 
%This is why N is usually considered a ``secondary'' element.
%If N were a purely secondary element and the stellar 
%lifetimes were neglected (Instantaneous Recycling Approximation),
%we should expect that 
%$(N/O)\propto(O/H)$. In reality, such a linear
%behaviour is never observed either in the
%SN nor in blue compact galaxies. In particular, 
%in the SN there seems to be a small plateau at low metallicity 
%then an almost linear increase followed by a flattening at high metallicities.
%The bulk of Nitrogen 
%probably originates from low and 
%intermediate 
%mass stars as both secondary and primary element.
%Primary N can, in fact, originate in AGB stars when the 
%hot-bottom burning acts in 
%conjunction with the third dredge-up (Renzini \& Voli, 1981;
%van den Hoek \& Groenewegen 1997) and also in massive stars as 
%a result of stellar rotation (Maeder\& Meynet 2001). However,
%the N produced by massive stars, either in a secondary or primary fashion, 
%is negligible.
%N is expected to be  
%restored into the ISM after oxygen, with a time delay depending
%either on nucleosynthesis or stellar lifetimes. 

N is expected to be  
restored into the ISM after oxygen, with a time delay depending
on the lifetimes of its stellar progenitors.
In particular, N is mainly produced in low and intermediate mass stars
which start dying after $\sim 3 \cdot 10^{7} yr$ since the beginning of 
star formation. The bulk of N production occurs after $\sim 250-300$ Myr
(Henry et al. 2000; Chiappini et al. 2002).
 For this reason the [N/O] ratio, provided that its evolution is computed correctly, 
is a useful ``cosmic clock'', in the sense 
that it can be important in dating
galaxies (Matteucci et al. 1997).
Therefore,  the N/O ratio in DLAs is important in order to 
understand the nature and the age of DLAs. 
(Pettini et al. 1995, Lu et al. 1998, Centuri\'on et al. 1998, 
Ellison et al. 2001). 
Matteucci et al. (1997) attempted to explain the N/O ratios observed in 
several DLAs by comparing them with the evolutionary tracks of [N/O]
computed for galaxies with different histories of star formation.
They concluded that some of the examined 
systems, those with the highest N/O, could be dwarf irregulars 
experiencing their first
or one of their first bursts of star formation, whereas others could
be spirals in the earliest phases of their evolution. They also studied 
the nature of the 
production of N as primary or secondary element, showing the possibility 
of explaining large N/O ratios even at low O abundances if
primary N were produced by massive stars.
In the present study we assume a set of yields for intermediate mass stars 
in which N is produced both as primary and secondary element 
(Van den Hoeck and Groenwegen 1997).
However, the calculations of the yields for N 
for low and intermediate mass stars 
present some uncertainties, since they depend on several free parameters,
as is discussed in Chiappini et al. (2002).
Owing to these uncertainties, all our models tend to overestimate the N abundances,
which should be regarded as upper limits.\\
The number of nitrogen measurements in DLAs is not very
large owing to the difficulty of measuring the NI lines inside
the Ly-$\alpha$ forest. In order to have a sample
of [N/$\alpha$] data sufficiently large,
we have chosen to study the [N/Si] ratio, since  it
is much easier to measure Si than O in DLAs.
Nitrogen is believed to be unaffected by dust, but
silicon experiences some 
depletion, which may affect the [N/Si] and [Si/H] measurements.
%%%%%%%%%%%%%%%%%%%%%%%%%%%%%%%%%%%%%%%%%%%%%%%%%%%%%%%%%%%%
%The sample of DLAs with [N/Si] data which can be corrected for
%dust includes only 5 systems, 4 of which show negligible
%Si corrections.
%%%%%%%%%%%%%%%%%%%%%%%%%%%%%%%%%%%%%%%%%%%%%%%%%%%%%%%%%%%%%%
However, given the small amount of Si depletion observed in many systems,
we have chosen to plot
in Figs. 12 and 13 the full sample of DLAs with
[N/Si] data not corrected for dust, which includes 17 systems.
For a few of them, the [N/Si] may be overestimated and the
[Si/H] underestimated owing to dust effects.\\
From Figure 12, 
one can see  that most of the systems lie at redshifts $z < 3$ indicating 
the possibility that they have formed at $z \sim 3$ rather than at $z =5$ as 
assumed in the figure. This fact reinforces the conclusion that it is problematic 
to assume that the abundances versus redshift can be interpreted as an 
evolutionary diagram in analogy with the stellar abundances in the Galaxy.
Moreover, Figure 12 suggests that the majority of DLAs  correspond either to outer regions 
of spirals or irregular galaxies.
In Figure 13 we report the [N/Si] versus [Si/H] where
it is evident that 
the inner regions of spirals and the solar neighbourhood region
have very similar and smooth curves, so the study of the N/Si ratio 
is not suited to distinguish
between these regions. 
On the other hand,
the model corresponding to the most external radius of the spiral galaxy shows a 
very different behaviour. 
This is again the effect due to 
the star formation threshold.
Again, from Figure 13 we can conclude that most of the observed 
DLA systems can be 
reproduced either by different regions of spiral disks or by dwarf irregular
starbursting galaxies (see also Chiappini et al. 2002).
In fact, in a regime of starburst the [N/Si] ratio is decreasing and increasing 
during the starburst and interburst phases,
respectively. We are not showing here all the possible cases to avoid complications, 
but we suggest  that 
objects with relatively 
high [Si/H] and undersolar [N/Si] could also be dwarfs observed during a starburst 
which follows 
previous star formation events.
The predictions for ellipticals are not shown in Figure 13 because they 
would lie at too high metallicities. This reinforces the conclusion that 
ellipticals are not likely to be DLA objects.
\\

%%%%%%%%%%%%%%%%%%%%%%%%%%%%%%%%%%%%%%%%%%%%%%%%%%%%%%%%%%%%%%%%%%%%%%%%%%%%%%%%%%%%%%%%%%%%%%%%%%%%%%%%%%%%%
\begin{figure*}
\centering
\vspace{0cm}
\psfig{file=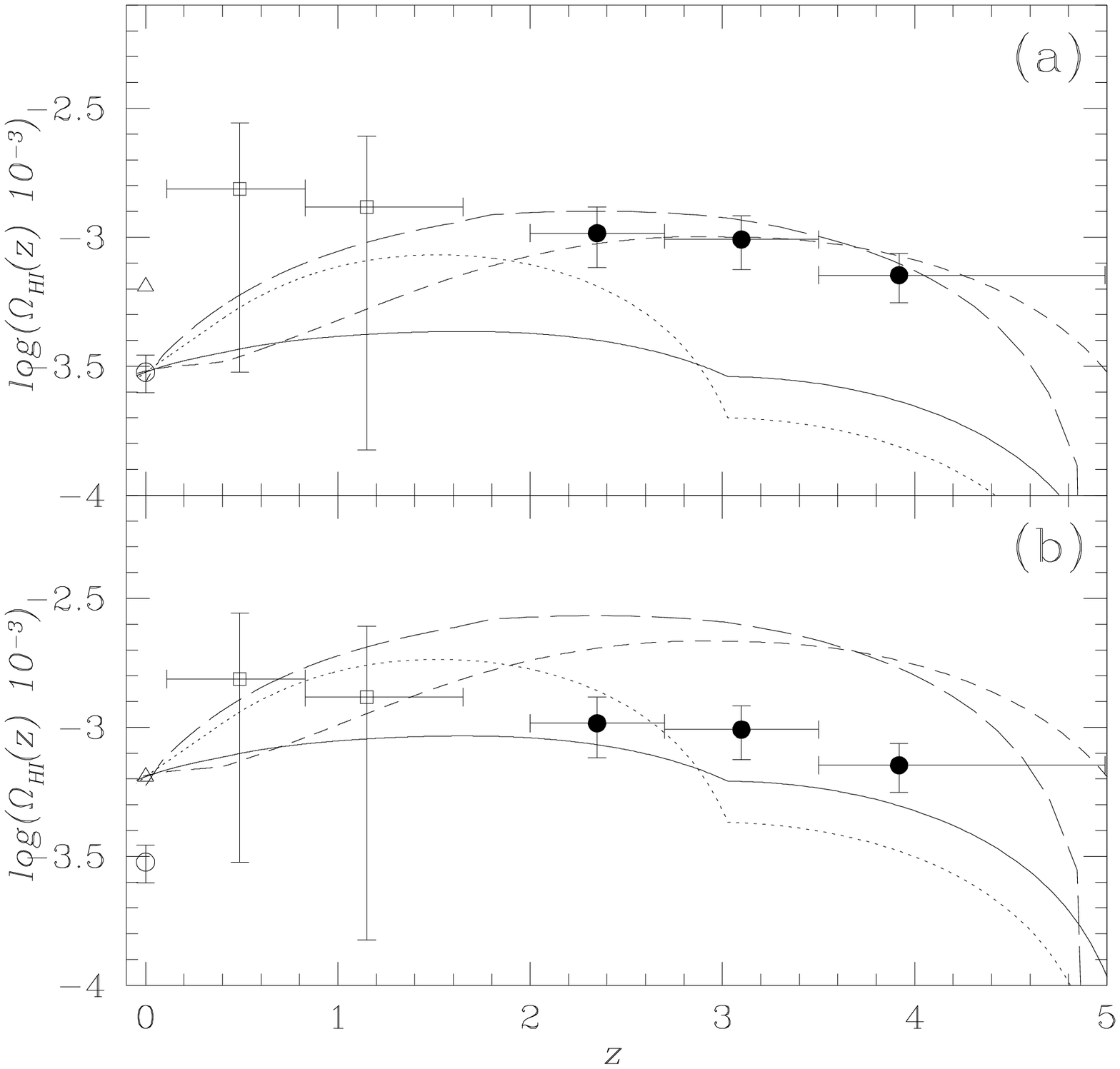 ,height=10cm,width=10cm}
\caption[]{Evolution of the comoving density of neutral gas with redshift in case of galaxy formation 
at $z_{f}=5$. 
In the upper panel the model curves have been normalized to the $z=0$ value by Zwaan et al. (1997),
whereas in the lower panel to the value by Rosenberg \& Schneider (2002).
 \emph{Solid line:} spiral galaxy at R=8 kpc;
 \emph{dotted line:} spiral galaxy at R=4 kpc;
 \emph{short-dashed line:} spiral galaxy at R=14 kpc;
 \emph{long-dashed line:} magellanic irregular galaxy (LMC).
\emph{Open circle}: value inferred from 21 cm emission from local galaxies (Zwaan et al. 1997);
\emph{Open triangle}: value by Rosenberg \& Schneider (2002); 
\emph{solid circles}: values by P\'eroux et al.  (2001);
\emph{open squares}: values by Rao and Turnshek (2000).
 
  }	
\end{figure*}
%%%%%%%%%%%%%%%%%%%%%%%%%%%%%%%%%%%%%%%%%%%%%%%%%%%%%%%%%%%%%%%%%%%%%%%%%%%%%%%%%%%%%%%%%%%%%%%%%%%%%%%%%%%%
\subsection{The evolution of the HI density}

The high redshift neutral gas density is traced mainly by DLAs, since 
these systems represent the largest gas reservoirs
and are likely to dominate the mass density of neutral gas in the universe 
(Storrie-Lombardi et al. 1996).
Recent DLA studies at low redshift ($z<2$, Rao \& Turnshek 2000) seem to indicate,
when compared with data at high redshift,
a weak
(or totally absent) evolution of  $\Omega_{HI}$
(i. e. the comoving HI density divided by the critical density of the universe),
with a nearly constant value.
On the other hand, the $z=0$ values inferred from the 21 cm emission studies (Zwaan et al.
1997) and optical surveys (Rao \& Briggs 1993) are noticeably lower than 
the high redshift estimates, thus indicating some evolution.
According to  these studies, large spiral galaxies
have been confirmed to be the major contributors to the total HI mass in the local universe
(see also Zwaan et al. 2001).
However, the very recent Arecibo Dual-Beam survey (Rosenberg \& Schneider 2002) yields a total HI mass 
density substantially higher than the previous estimates, with a non-negligible contribution 
from low-mass sources.
In summary, the situation is still unclear and it is difficult to assess from the available
data whether there is evolution or not.
In Figure 14 we plot the evolution of the comoving density of neutral gas 
with redshift as predicted for irregular and spiral galaxies at different radii
 and compared with various observations (P\'eroux et al. 2001, 
Rao \& Turnshek 2000, Zwaan et al. 1997, Rosenberg \& Schneider 2002).
In figures 14 (a) and (b)
the predicted curves have been normalized in order to reproduce the $z=0$ value 
measured by  Zwaan et al. (1997) and Rosenberg \& Schneider (2002), respectively.
In the case of the upper panel, where we consider the lower value for $\Omega_{HI}$
 at the present time (Zwaan et al. 1997),
both the external disk (R=14 kpc) and the irregular galaxy seem the most promising in reproducing the 
data, since they represent the slowest evolving systems having
at any time (i.e. redshift) the right amount of gas to reproduce each point.\\ 
In the case of the lower panel (figure 14b), where the higher value from Rosenberg \& Schneider (2002) is adopted
for the normalization at $z=0$,
the irregular and external spiral fail in reproducing the high redshift
points, which are substantially lower than the predictions, whereas the solar neighbourhood curve is
the one in best agreement with each point.
%%%%%%%%%%%%%%%%%%%%%%%%%%%%%%%%%%%%%%%%%%%%
%In the previous sections we have seen that 
%the SN/inner disk had a too strong evolution to reproduce the abundance patterns for several elements 
%([Z/H], [N/Si]), but they were in reasonable agreement with the dust-corrected [Si/Fe] and [S/Zn] values
%observed in the most metal rich DLA systems.
%%%%%%%%%%%%%%%%%%%%%%%%%%%%%%%%%%%%%%%%%%%%
We conclude that the study of the neutral H evolution cannot be considered as a good diagnostic in
order to understand which galaxies are dominant in the high-redshift DLA population, 
at least until the low redshift measures will be more homogeneous than they are at the present time.
%%%%%%%%%%%%%%%%%%%%%%%%%%%%%%%%%%%%%%%%%%%
%Most observations support the argument that the slow evolution 
%of both neutral gas and metallicity indicates that DLA systems
%trace a slowly evolving population of objects (Rao and Turnshek 2000, 
%Pettini et al. 1999, Jimenez et al. 1999). Therefore, they cannot
%represent the main contributors to the strong evolution in the cosmic star 
%formation rate observed at low (i. e. $<1$) redshifts 
%(see e. g. Steidel et al. 1999, Rowan-Robinson 1999, Madau et al. 1996), 
%probably due to various biases which exclude the most dense, metal
%rich systems (Boiss\'e et al. 1998, Pettini et al. 1999).
%%%%%%%%%%%%%%%%%%%%%%%%%%%%%%%%%%%%%%%%%%%
In fact, from the observational point of view, 
the low 
redshift region ($z<1.6$) is particularly uncertain, since 
the Lyman-$\alpha$ line drops out of the visible atmospheric window
and many
systems are supposed to be metal (i. e. dust)-rich causing severe 
complications in the measurements. 
A larger sample of DLAs,
preferentially based on a radio-selected QSO sample,
is therefore required 
in order to shed light on the real evolution of the neutral gas in the redshift range 
untestable with the 
earth based optical telescopes.
The first results in this field come from the CORAL survey (Ellison et al. 2001a), in which a set 
of 22 DLAs has been assembled.
Although the authors stress that their results are tentative, 
given the limited size of the data set, they conclude that 
magnitude limited, $z=2-3.5$ surveys, based on optically selected QSOs, 
underestimate the number of DLAs and the detected neutral gas mass by no 
more than a factor of two.\\

\section{Conclusions}
We have investigated the chemical properties of DLA systems by means of 
detailed chemical evolution models for galaxies of different
morphological types, i. e. ellipticals, irregulars and spirals. 
In the case of spiral galaxies, we run a multizone code in order to
disentangle the different chemical evolution histories at various 
galactocentric radii, which can differ substantially from each other,
giving rise to very
different abundance patterns. We compared our predictions 
for  [Si/Fe], [Ni/Fe], [Zn/Fe] and [N/Si]
with observational data either as functions of [Fe/H] or redshift.  
The simultaneous comparison of the abundance ratios 
as functions of metallicity (e.g. [Fe/H]
and/or [Zn/H]) and redshift provides
stronger constraints on the nature of DLA systems, and this 
is the first time that this possibility was taken into account.
Our main goal was to infer the nature and possibly the age
of the DLA population, 
by means of the [$\alpha$/Fe] and the [N/$\alpha$] ratios. 
We have assumed that all galaxies formed either at redshift 
$z_{f}=3$ or $z_{f}=5$, but this assumption, which might be unrealistic,  
influences only the diagrams 
as functions of redshift, not those as functions of metallicity.
In several cases, we have compared our model predictions 
with
abundance 
measurements not dust-corrected with the exception of the [Si/Fe] and [Ni/Fe]
ratios versus [Fe/H] which have been
compared also with a set of data dust-corrected according to 
the prescriptions of Vladilo (2000a,b). 
Finally, we have studied the redshift evolution of the neutral 
gas density, which can be useful to gain information
on the formation epoch and global evolution of the DLA population 
and on the average evolution of the neutral gas in the universe, 
provided that
DLAs really represent the largest gas reservoirs, i. e. that biases 
do not affect the observations, thus
excluding some systems from the samples.\\
Our main results can be summarized as follows:\\
1) Under the assumption that galaxy formation occurred at $z_f=5$ for all the objects,
we can reproduce the observed dispersion in DLA metallicity 
in the plot [Zn/H] versus redshift, by considering 
evolutionary models for disk galaxies computed at different distances
from the galactic centre. 
On the other hand, we cannot exclude that the same DLAs have started forming stars at different cosmic epochs.
Therefore, DLA systems cannot be treated as the stars in the Milky Way 
which trace, with their metallicity, the evolutionary history of the 
gas in the Galaxy.  
Our results, together with
observational evidence concerning DLA systems, namely:
i) at $z\sim2$ the typical metallicity of the DLA galaxies is
approximately $1/10$ of solar, ii) the 
paucity or absence of low ($<100$ solar) 
metallicity objects and iii) a relevant metal enrichment observed in many
systems at $z\sim3$, leads us to suggest that a substantial fraction of 
the DLA 
systems was already in place at $z>3$ and that is likely that they started forming stars between z=5 and z=3.\\
2) The $\alpha$-enhancements relative to refractory elements (e.g. Si and Fe)
observed in many systems 
vanishes once dust-corrections are considered. The resulting almost 
solar values for these ratios
are found to be consistent with the 
evolution of both irregular (including star-bursting dwarfs) galaxies or spiral disks 
seen at various galactocentric distances.
For the DLAs with the lowest metallicities the irregulars and the outer regions
of spiral disks 
best reproduce the data. In fact, the external regions of spiral disks,  
where the low density causes frequent oscillations in the star 
formation above and below the threshold level, show a behaviour similar to
starburst galaxies. Such a weak and gasping star formation history seems
to be 
the most promising in reproducing the majority of the observations ([$\alpha$/Fe], [N/$\alpha$], [Zn/H], [Ni/Fe]). 
This regime of star formation is also typical of dwarf irregulars and 
low surface brightness galaxies and therefore we cannot exclude that 
DLAs could also be such systems.
However, both irregulars and outer disks cannot cover all the metallicity 
range spanned by the observations.
At higher metallicities the observations are better
reproduced by considering an inner disk/solar
neighbourhood-like evolution.
This result is confirmed by the comparison between model predictions 
and observations of the [S/Zn] ratios, both of which are
unaffected by dust depletion, so their pattern reflects pure 
nucleosynthesis.

By means of a photometric code (Jimenez et al. 1998) matched to our
chemical evolution models,
we have performed 
a prediction regarding the luminosity of a DLA system either if the absorption occurred 
in a spiral or an irregular galaxy.
Assuming galaxy formation at $z_{f}=5$, at $z\sim4$
the magnitude of a typical spiral in the 
I band is $m_{I}= -20.346$, which corresponds to a luminosity $L/L_{LBG}\sim 0.09$, 
where $L_{LBG}$ is the charactistic luminosity of the Lyman Break Galaxies at $z\sim4$
(Steidel et al. 1999).
In the case of an irregular galaxy, $m_{I}= -16.48$, corresponding to a luminosity 
$L/L_{LBG}\sim 0.003$.
Both values are consistent with the limits of $L/L_{LBG}< 0.11$ derived 
by Prochaska et al. (2002a) for DLA systems.\\
3) The predictions of our models indicate that, at least to first order,
the evolutions of Fe and Zn can be considered strictly linked and 
independent of metallicity.
Based on this result, the observed spread in the 
[Zn/Fe] vs [Fe/H] diagrams 
should be due more to the dust 
(like differential dust depletion of Fe and Zn, differential dust 
pollution along different lines-of-sight, different amounts of dust
within DLA systems) than to pure nucleosynthesis. We do not feel safe in 
drawing conclusions about the behaviour of the [Ni/Fe] ratios because of the 
uncertainties still present in the nucleosynthesis of nickel.\\
4) Given the uncertainties in low redshift measures, at the present time
the study of the neutral H evolution cannot be considered as a good diagnostic in
order to understand which galaxies are dominant in the high-redshift DLA population.
A larger sample of DLAs,
preferentially based on a radio-selected QSO sample,
is required 
in order to shed light on the real evolution of the neutral gas in the redshift range 
untestable with the 
earth based optical telescopes.\\
5) From the chemical evolution point of view,
elliptical galaxies are the most unlikely DLA candidates at 
all metallicities because of the very high metallicities and
overabundances of $\alpha$-elements reached at the early phases of 
their history, when a type II SN chemical enrichment is dominant.
Elliptical galaxies are perhaps better candidates to explain the Lyman-break 
galaxies (see Matteucci \& Pipino, 2002).

\section*{Acknowledgments}
We are grateful to I.J. Danziger for reading the manuscript and useful 
suggestions.
We also thank C. Chiappini, C. P\'eroux, A. Pipino
and S. Cristiani
for many useful discussions, and
H. Umeda for having kindly provided the compilation of data in figure 2.
Finally, we wish to thank an anonymous referee for many comments which improved
 the quality of our work.

This paper has been typeset from a \LaTeX\ file prepared by the author.

%\appendix

%\section[]{}

%()

%\subsection{Subsection title}

%\bsp

\label{lastpage}

\end{document}